\documentclass[%
 reprint,
 amsmath,amssymb,
 aps,
prb,
]{revtex4-2}

\usepackage{graphicx}
\usepackage{dcolumn}
\usepackage{bm}
\usepackage{hyperref}
\usepackage{color}
\usepackage{mathtools}
\usepackage{hyperref}

\hypersetup{colorlinks=true, linkcolor=blue, urlcolor=blue}

\begin{document}

\title{Envelope-function theory of inhomogeneous strain in semiconductor nanostructures} 

\author{Andrea Secchi}
\email{andrea.secchi@nano.cnr.it} 
\affiliation{Centro S3, CNR-Istituto di Nanoscienze, I-41125 Modena, Italy} 
\author{Filippo Troiani}
\affiliation{Centro S3, CNR-Istituto di Nanoscienze, I-41125 Modena, Italy} 

\date{\today}

\begin{abstract}
Strain represents an ubiquitous feature in semiconductor heterostructures, and can be engineered by different means in order to improve the properties of various devices, including advanced MOSFETs and spin-based qubits. However, its treatment within the envelope function framework is well established only for the homogeneous case, thanks to the theory of Bir and Pikus. Here, we generalize such theory to the case of inhomogeneous strain. By fully accounting for the relativistic effects and metric aspects of the problem, we derive a complete envelope-function Hamiltonian, including the terms that depend on first and second spatial derivatives of the strain tensor. 
\end{abstract}
 
\maketitle

\section{Introduction}

Strain represents a common feature in semiconductor nanostructures. It develops spontaneously during their fabrication process, because of the lattice mismatch between heterogeneous layers, and can be induced by cooling the system to cryogenic temperatures, due to the presence of materials with different thermal-expansion coefficients \cite{Pla18a, Lo15a,Thorbeck15a}. As an uncontrolled or unaccounted phenomenon, strain can result in significant differences between the actual and the nominal properties of the nanostructure. On the other hand, strain can be intentionally engineered, in order to modulate the band structure and increase the carrier mobility, an approach that is actively pursued, e.g., with MOSFETs \cite{Sverdlov, Sun07a} or silicon nanowires \cite{Niquet12a}. 

These effects are particularly relevant in semiconductor-based implementations of quantum computing. Silicon and germanium quantum dots have emerged as promising hosts of electron- or hole-spin qubits \cite{Zwanenburg13a, Yoneda18a, Scappucci21a, Chatterjee21a, Stano22a, Zwerver22a, Saraiva22a, Burkard23a, Venitucci18a, Venitucci19a, Hetenyi20, Bosco21a, Bosco21b, Bosco21c, Secchi21a, Bellentani21a, Secchi21b, Secchi23a, Secchi23b, Forghieri23a, Fernandez22a}, whose properties can be strongly affected by strain. In particular, it has been shown that in these systems inhomogeneous strain can modify both the localization of the confined particle and its coupling to external fields, specifically through a modulation of the Rabi frequency \cite{Abadillo23a} and of the $g$-factor \cite{Liles21a}. 

The tool of election for simulating the properties of spin qubits in semiconductor quantum dots is represented by the Luttinger and Kohn (LK)'s envelope-function formalism \cite{Luttinger55a, VoonWillatzen, Winkler}. This applies to crystalline systems subjected to a spatially slow-varying external potential, such as the one generated by the metal gates used in electrostatically defined nanostructures. Describing the effects of strain on the electron and hole states requires an extension of LK's theory, which was developed by Bir and Pikus (BP) for the case where the strain tensor is small and homogeneous \cite{BirPikus, Richard03a, Chao92a, Bahder90, Bahder92}. Even in these conditions, the absolute displacements of the ions (with respect to the unstrained crystal) may be comparable or larger than the lattice constant. This makes the displacements unsuitable as an expansion parameter for the electron-nuclei potential, unlike for the theory of electron-phonon interactions \cite{Whitfield61}. BP's key idea was to introduce a new set of electron coordinates that make the power expansion of the electron-nuclei potential in the strain tensor possible, thus enabling a perturbative calculation of the electron and hole states. 

In view of the above, a generalization of BP's theory to the case of inhomogeneous strain would be highly desirable, but is far from trivial. In our understanding, the previous attempts that have been made in this direction are affected by significant shortcomings. These consist either in an incorrect treatment of the Schr\"odinger equation in the required set of curvilinear coordinates, resulting in the non-hermiticity of the particle Hamiltonian \cite{Zhang94a}, or in the use of a non-practical basis set within a non-relativistic treatment, which precludes from the outset an accurate description of spin-orbit interactions \cite{Li14a}.

In this Article, we extend BP's theory to the case of inhomogeneous strain in a rigorous and comprehensive way. This is achieved by properly taking into account the modifications to the quantum-mechanical formalism that arise when the metric is non-Cartesian \cite{DeWitt52a, Gneiting13a}, and by including relativistic corrections to the Schr\"odinger equation via a low-energy expansion of the covariant Dirac equation \cite{Breev16a}. Our central result --- applicable to a variety of semiconductor nanostructures, in the presence of slowly-varying inhomogeneous strain and external electrostatic potential --- is a set of equations, whose solution gives the envelope functions within a manifold of arbitrary dimension. From these we derive, as a relevant case, the strain-related 6-band Hamiltonian for the hole states in silicon and germanium, and more generally in crystals with diamond structure. For the sake of readability, the main logical steps that have been followed are reflected in the structure of the main text, which contains the main results. The complete derivations are reported in the Appendices and in the Supplementary Material (SM) \cite{SuppMat}, to which we provide detailed reference at each step.

\section{Inhomogeneous strain}

The first step consists in the introduction of a curvilinear coordinate system, which allows to express the nuclei potential in the strained system as a perturbative expansion in the strain tensor. This approach, introduced by BP for the case of homogeneous strain and generalized to the inhomogeneous case by Zhang \cite{Zhang94a}, is recalled here for the reader's convenience.
In the original Cartesian reference frame, let $\boldsymbol{r}_{\rm C}$ define the electronic coordinates, while $ \boldsymbol{R}_{i,0} $ and $ \boldsymbol{R}_{i} \equiv \boldsymbol{R}_{i,0}+\boldsymbol{u}_i $ are the nuclei positions in the absence and in the presence of strain, respectively. Generalizing BP's approach to the case of inhomogeneous strain, one introduces a set of curvilinear coordinates $\boldsymbol{r}$ to describe the electronic position. These are related to the $\boldsymbol{r}_{\rm C}$ by the equation \cite{Zhang94a}:  
\begin{align}
    r_{\rm C}^{\alpha} = r^{\alpha} + u^{\alpha}(\boldsymbol{r})   \,,
    \label{Jacobian}
\end{align}
where the Greek indices label vector components ($\alpha=1,2,3$). The continuous inhomogeneous displacement $\boldsymbol{u}(\boldsymbol{r})$ is assumed to be an invertible and differentiable function of $\boldsymbol{r}$ to all needed orders. It fulfils the conditions $\boldsymbol{u}(\boldsymbol{R}_{i,0}) = \boldsymbol{u}_i$ and $\left| \boldsymbol{u}(\boldsymbol{r}) \right| \ll \left| \boldsymbol{r} \right|$. 
The former condition allows for the expansion of the nuclei potential in powers of the strain in the transformed coordinate frame, while the latter condition follows from the assumption that the strain tensor is small everywhere.

Given the displacement functions $u^\alpha$, the components of the strain tensor can be defined as follows:
\begin{align}
      \varepsilon^{\alpha}_{\beta}(\boldsymbol{r}) \equiv \partial_\beta u^\alpha(\boldsymbol{r})\,,
    \label{definition of strain}
\end{align} 
where $\partial_\beta\equiv \partial/\partial r^\beta$.
Provided that the strain tensor is small and varies slowly over the scale of a unit cell, by applying the transformation in Eq.~\eqref{Jacobian}, one can express the potential $U_{\rm n}$ generated by the nuclei in the strained system in the form: 
\begin{align}
       U_{\rm n}(\boldsymbol{r})  \approx   U_{{\rm n},0}( \boldsymbol{r})  +  \varepsilon^{\alpha}_{\beta}\left( \boldsymbol{r} \right)  \, U^{\beta}_{\alpha}(\boldsymbol{r} )  \,.
      \label{expansion of EN potential}
\end{align}
Here, $U_{{\rm n},0}$ is the potential in the unstrained system and $U^{\beta}_{\alpha}$ is a strain-independent function that has the same periodicity as the unstrained lattice, while the product $\varepsilon^{\alpha}_{\beta}\left( \boldsymbol{r} \right)  \, U^{\beta}_{\alpha}(\boldsymbol{r} )$ is in general not periodic. Further details on the nuclei potential relations are provided in Appendix \ref{app: preliminary}. In Appendix \ref{app: point transf}, instead, we give a formal derivation of the relations between Hamiltonians and wave functions in two coordinate systems connected by a point transformation such as Eq.~\eqref{Jacobian}. Here and in the remainder of this Article, we use Einstein's summation convention on repeated Greek indexes.

\section{The Schr\"odinger problem in curvilinear coordinates}

The second step consists in deriving the general expression of the Schr\"odinger equation for an electron in curvilinear coordinates, with the inclusion of the spin-orbit term. The adoption of the curvilinear coordinates $\boldsymbol{r}$ implies the introduction of a nontrivial metric tensor, i.e. a $g_{\mu \nu} \neq - \delta_{\mu \nu}$ \cite{DeWitt52a}. As a result, the matrix element of a local operator $\hat{A}$ between two arbitrary electron (spinorial) states is given by:
\begin{align}
    \big< \Psi \big| \hat{A} \big| \Phi \big> = \int d \boldsymbol{r}\, \sqrt{-g(\boldsymbol{r})}\, \Psi^{\dagger}( \boldsymbol{r} )\, \cdot A(\boldsymbol{r}) \, \Phi(\boldsymbol{r}) \,,
    \label{scalar product in arbitrary metric}
\end{align}
where $g(\boldsymbol{r}) =   \det \left[ g_{\mu \nu}(\boldsymbol{r}) \right]$. The definition of inner products can be obtained from the above equation simply by replacing the generic operator $A$ with the identity. As a technical but crucial point, we note that, in a curvilinear coordinate system, the definition in Eq.~\eqref{scalar product in arbitrary metric} should be used in evaluating the hermiticity of operators and the scalar products between states, rather than its Cartesian counterpart, corresponding to $\sqrt{-g(\boldsymbol{r})} = 1$ \cite{Zhang94a}. 

In order to obtain the correct Hamiltonian in curvilinear coordinates and to include spin-orbit coupling, we generalize the covariant formulation of the Schr\"odinger equation given in Ref.~\cite{DeWitt52a}, which applies to a non-relativistic Hamiltonian. Starting from the covariant Dirac equation for the 4-component electron field in an electromagnetic potential \cite{Breev16a}, which holds for arbitrary metric tensors, we take the non-relativistic limit and allow for a nontrivial metric in the spatial sector only. The result is a Schr\"odinger equation for 2-spinors that can be augmented with any order of relativistic corrections, while inheriting the covariance of the initial Dirac equation. 

In the absence of magnetic field and up to the first order in the relativistic corrections, the Hamiltonian can be written as: $H = H_{\rm kin} + H_{\rm rel} + U$, where the kinetic term reads
\begin{align}
      H_{\rm kin}    =    - \frac{  \hbar^2 }{2 m} \left[\left( \nabla_{\rm C}^2 r^{\nu} \right)   \partial_{\nu} - g^{\mu \nu} \partial_{\mu}      \partial_{\nu} \right] \,,
      \label{Schroedinger curvilinear, no A}
\end{align} 
$U$ is a generic scalar potential, and the dominant component in the relativistic term is given by the spin-orbit Hamiltonian
\begin{align}
      H_{\rm so} 
      & = - \frac{ {\rm i}  \hbar^2   }{4 m^2 c^2 }    \left(    \frac{\partial r^{\mu}}{\partial r^{\alpha}_{\rm C}} \frac{\partial r^{\nu}}{\partial r^{\beta}_{\rm C}} \sigma^{\alpha \beta} \right) \left( \partial_{\mu} U \right)  \partial_{\nu}  \,.
      \label{rel corrections curvilinear, no A}
\end{align} 
Here, we adopt the notation $\nabla^2_{\rm C} \equiv \partial^2 / (\partial r^{\alpha}_{\rm C} \partial r^{\alpha}_{\rm C})$. Besides, $\sigma^{\alpha \beta} = -\widetilde{\epsilon}^{\, \alpha \beta \gamma} \sigma_{\gamma}$, where $\widetilde{\epsilon}^{\, \alpha \beta \gamma}$ is the invariant Levi-Civita symbol and $-\sigma_{\gamma}$ are the Pauli matrices.

Relying on the generalized expression of the matrix elements and of the inner product [Eq.~\eqref{scalar product in arbitrary metric}], one can write the matrix elements of the Hamiltonian between 2-component spinors in a manifestly Hermitian way, assuming that the wave functions either vanish at infinity, or satisfy the Born-von Karman boundary conditions. Further details on the relativistic terms of the Hamiltonian and on the boundary conditions are provided in Appendices \ref{app: derivation Schr} and \ref{app: manifestly}, respectively.  

\section{Curvilinear coordinates from the strain tensor}

The equations reported in the previous paragraph provide a general framework, which can be applied to the present problem, where the non-Cartesian character of the coordinates results from the presence of inhomogeneous strain. 
In fact the strain tensor determines the Jacobian 
$J^\alpha_\beta\equiv\partial_\beta\partial r_{\rm C}^{\alpha}=\delta^\alpha_\beta+\varepsilon^\alpha_\beta$, as can be deduced from Eqs.~\eqref{Jacobian} and \eqref{definition of strain}. 
As to the metric tensor, to first order in the strain tensor, it can be expressed as 
\begin{align}
    & g^{\mu \nu}   \approx  -\delta^{\mu \nu}   -  \varepsilon^{\nu}_{\mu}   -     \varepsilon^{\mu}_{\nu}  \,,  
\end{align}
under the assumption that $\lVert \varepsilon_i(\boldsymbol{r}) \rVert \,\ll 1 $, and thus $J^{-1} \approx 1 - \varepsilon$. From this it also follows that $\sqrt{- g } \approx 1 + {\rm tr}(\varepsilon)$.

The above relations provide the dependence of $H_{\rm kin}$ and $H_{\rm so}$ on the strain tensor, mediated by the inverse Jacobian, $\nabla_{\rm C}^2$ and $g^{\mu\nu}$.
To the first order in the strain tensor, the kinetic and spin-orbit components of the Hamiltonian can thus be written as follows:
\begin{align}
   &  H_{\rm kin}   = - \frac{  \hbar^2 }{2 m}   \left[  \frac{\partial^2}{\partial r^{\mu} \partial r^{\mu}}  + \overleftarrow{\partial}_{\mu} \left(   \varepsilon^{\nu}_{\mu}  + \varepsilon^{\mu}_{\nu}     
 \right)  \overrightarrow{\partial}_{\nu}   \right]   \,, 
    \label{Schroedinger Hamiltonian kin}
\\   
& H_{\rm so}   = -   {\rm i} \eta          \Big(         f_{\mu}  \sigma^{\mu \nu}     \overrightarrow{\partial}_{\nu}     -    \overleftarrow{\partial}_{\nu}     \sigma^{\mu \nu}      f_{\mu} +  \Sigma^{\nu} \overrightarrow{\partial}_{\nu}     -    \overleftarrow{\partial}_{\nu}    \Sigma^{\nu} \Big)        \,,
    \label{Schroedinger Hamiltonian so}
\end{align}
where
\begin{align}
    & \eta\equiv\hbar^2 / (8 m^2 c^2 ) \,, \nonumber \\
    & f_\mu \equiv\partial_\mu U_{{\rm n}, 0} \,, \nonumber \\
    & \Sigma^{\nu}    \equiv  \left( \partial_{\mu} \varepsilon^{\alpha}_{\beta}  \right) U^{\beta}_{\alpha}  \sigma^{\mu \nu} +  \varepsilon^{\alpha}_{\beta}    \left( \partial_{\mu}  U^{\beta}_{\alpha}  \right)  \sigma^{\mu \nu}    -    \varepsilon^{\nu}_{\beta}        f_{\alpha}  \sigma^{\alpha \beta}  \nonumber \\
    & \quad \quad 
    -   \varepsilon^{\alpha}_{\mu}     f_{\alpha} \sigma^{\mu \nu} \,,
    \label{Sigma nu}
\end{align} 
and the arrows above the differential operators specify whether these must be applied to the wave function on the left or right sides of the Hamiltonian when evaluating its matrix elements.

As to the potential, induced by the nuclei, its expression is given by the sum of the unstrained contribution and of a perturbation that depends linearly on the strain tensor [Eq.~\eqref{expansion of EN potential}].

The derivations of the above equations can be found in Appendices \ref{app: derivation Schr} and \ref{app: manifestly}.

\section{Generalized Luttinger-Kohn theory}

In the LK solution scheme, the electron Hamiltonian matrix is derived in an orthonormal basis, and then reduced to a block structure by means of a suitable canonical transformation, which effectively separates the relevant manifold from the others, while perturbatively accounting for the inter-manifold coupling. In the present Section, this procedure is generalized in order to include the case of a curvilinear set of coordinates, with consistently defined orthonormality relations.

In order to identify a complete basis set, we initially consider the part of the Hamiltonian that is of order zero in the strain. Being this a periodic function of $\boldsymbol{r}$, one can apply Bloch's theorem in order to derive its eigenfunctions $ \psi_{n, \boldsymbol{k}} = {\rm e}^{{\rm i} \boldsymbol{k} \cdot \boldsymbol{r}}  u_{n, \boldsymbol{k}}(\boldsymbol{r}) $ and eigenvalues $E_{n}(\boldsymbol{k})$. 
In view of an expansion around, e.g., the ${\bf\Gamma}$ point, it is convenient to introduce also the LK functions \cite{Luttinger55a} $\chi_{n, \boldsymbol{k}} \equiv {\rm e}^{{\rm i} \boldsymbol{k} \cdot \boldsymbol{r}} u_{n, \boldsymbol{0}}(\boldsymbol{r})$. In the curvilinear coordinates neither the Bloch nor the LK functions form an orthonormal basis \cite{letmeexplain}, according to the inner product defined in Eq.~\eqref{scalar product in arbitrary metric}. 
However, the orthonormality relations can be recovered by suitably modifying the LK functions, according to the relations \cite{Whitfield61}: 
\begin{align} 
    & \overline{\chi}_{n, \boldsymbol{k}} \equiv \frac{  \chi_{n, \boldsymbol{k}}}{  \left[-g(\boldsymbol{r}) \right]^{1/4}}  \approx \left[ 1 - \frac{1}{2} {\rm tr}\,\varepsilon(\boldsymbol{r})  \right] {\rm e}^{{\rm i} \boldsymbol{k} \cdot \boldsymbol{r}} u_{n, \boldsymbol{0}}(\boldsymbol{r}) \,.
     \label{orthonormal set primed coordinates}
\end{align}
Analogous modifications can be applied in order to recover the orthonormality relations for the Bloch functions.

In the modified LK basis, the matrix elements of the first-order strain-dependent component of the Hamiltonian read:
\begin{widetext}
\begin{align}
       \big< \overline{\chi}_{n, \boldsymbol{k}} \big|      \hat{H}^{(1)}        \big| \overline{\chi}_{n', \boldsymbol{k}'} \big>     =  - \frac{  \hbar^2  }{4 m}       \left| \boldsymbol{k} - \boldsymbol{k}' \right|^2         \widetilde{\varepsilon}^{\, \mu}_{\mu}(\boldsymbol{k} - \boldsymbol{k}') \, \delta_{n,n'} 
 + \widetilde{\varepsilon}^{\, \mu}_{\nu}(\boldsymbol{k} - \boldsymbol{k}')    \left(  \mathcal{D}_{\mu}^{\nu}        + k^{\alpha}  \mathcal{L}_{\alpha; \mu}^{\nu}   
+ k'^{\alpha} \mathcal{L}_{\alpha; \mu}^{*\nu} + k^{\alpha} k'^{\beta}  \mathcal{Q}_{\alpha \beta; \mu}^{\nu} \right)_{n, n'}       \,,
    \label{matrix element LK}
\end{align}
\end{widetext}
where $\widetilde{\varepsilon}^{\, \mu}_{\nu}(\boldsymbol{q})$ is the Fourier transform of the strain tensor. The last parentheses on the right include the deformation-potential terms; in particular, the $\boldsymbol{k}$- independent quantities $\mathcal{D}$, $\mathcal{L}$ and $\mathcal{Q}$ are given respectively by  
\begin{subequations}\label{total deformation potentials}
\begin{align}
    & \mathcal{D}^{\nu}_{\mu} \equiv D^{\nu}_{\mu} + \Delta D^{\nu}_{\mu} \,, \\ 
    & \mathcal{L}_{\alpha; \mu}^{\nu} \equiv L_{\alpha; \mu}^{\nu} + \Delta L_{\alpha; \mu}^{\nu} \,, \\
    & \mathcal{Q}_{\alpha \beta; \mu}^{\nu} \equiv Q_{\alpha \beta; \mu}^{\nu} + \Delta Q_{\alpha \beta; \mu}^{\nu} \,,  
\end{align}
\end{subequations}
where the (dominant) non-relativistic components are 
\begin{subequations}\label{deformation potentials}
\begin{align}
    & D^{\nu}_{\mu} \equiv  U_{\mu}^{\nu}   -     \frac{1}{ m}   p_{ \mu} p_{\nu}  \,,   \\ 
    & L_{\alpha; \mu}^{\nu} \equiv - \frac{\hbar     }{2 m} \left(  \delta^{\nu}_{\alpha}  p_{\mu}   +   \delta^{\mu}_{\alpha}    p_{\nu}  \right) \,, \\ 
    & Q_{\alpha \beta; \mu}^{\nu}  \equiv - \frac{\hbar^2  }{2 m}  \left(  \delta^{\mu}_{\alpha} \delta^{\nu}_{\beta} 
+  \delta^{\nu}_{\alpha} \delta^{\mu}_{\beta} \right) \,,
\end{align}
\end{subequations} 
while the relativistic corrections read
\begin{subequations}\label{relativistic deformation potentials}
\begin{align}
& \Delta D^{\nu}_{\mu}   
      \equiv     {\rm i} \eta       \Big(    2 \overleftarrow{\partial}_{\alpha}    \sigma^{\alpha \beta} U^{\nu}_{\mu}   \overrightarrow{\partial}_{\beta} + \overleftarrow{\partial}_{\beta}   \sigma^{\nu \mu} f_{\beta}  \nonumber \\ 
   & \quad \quad \quad - \overleftarrow{\partial}_{\beta}  \sigma^{\nu \beta} f_{\mu}   -     \sigma^{\nu \mu} f_{\beta}   \overrightarrow{\partial}_{\beta}   
  +    \sigma^{\nu \beta} f_{\mu}   \overrightarrow{\partial}_{\beta}   \Big) \,, \\ 
& \Delta L_{\alpha; \mu}^{\nu}    \equiv   \eta    
     \Big(         2    \sigma^{\alpha \beta} U^{\nu}_{\mu}   \overrightarrow{\partial}_{\beta}       
     +  \sigma^{\nu \mu} f_{\alpha} 
     -  \sigma^{\nu \alpha} f_{\mu}   \Big)     \,, \\
& \Delta Q_{\alpha \beta; \mu}^{\nu}    \equiv      {\rm i} \eta \sigma^{\alpha \beta}    U^{\nu}_{\mu}     \,,
\end{align}
\end{subequations}
in terms of the quantities defined in Eqs.~\eqref{Sigma nu}. In the above, $p_{\mu} = - {\rm i} \hbar \partial_{\mu}$, and the quantities depending on the band indices $n$ and $n'$ in Eq.~\eqref{matrix element LK} are defined by the relation 
\begin{align}
    \left( A \right)_{n, n'} \equiv \frac{(2 \pi)^3}{\Omega_{\rm cry}} \int d \boldsymbol{r} \, u^{\dagger}_{n, \boldsymbol{0}}(\boldsymbol{r}) \cdot A\, u_{n', \boldsymbol{0}}(\boldsymbol{r}) \,.
    \label{matrix element nn'}
\end{align}

The next step consists in decoupling the low-energy manifold of interest ($n \leq N$) from the higher-energy states ($n > N$), using L\"owdin partitioning \cite{Lowdin51a, Winkler, VoonWillatzen, Zhang94a}. This amounts to applying a canonical transformation to the Hamiltonian, $\hat{\mathcal{H}} = e^{-\hat{S}}\, \hat{H}\, e^{\hat{S}} \equiv \hat{H} + \Delta \hat{H}$, and to its eigenstates, $|\phi\rangle = e^{-\hat{S}}\,|\psi\rangle$. The transformation is such that $\hat{\mathcal{H}}$ is approximately block-diagonal in band space, and specifically displays negligible coupling terms between the relevant manifold and the remote bands. 
If inter-manifold couplings related to the deformation-potential terms can be neglected, then the correction $\Delta \hat{H}$ for the $N$-dimensional low-energy manifold has the standard effective-mass form \cite{VoonWillatzen}.

Further technical details on the derivation of Eq.~\eqref{matrix element LK} are given in Appendix \ref{app: secular} and Section I of the SM \cite{SuppMat}. Details on the manifold decoupling are presented in Section II of the SM \cite{SuppMat}.

\section{Envelope functions}

The last step consists in the derivation of the confined particle states within the relevant $N$-dimensional manifold. The electron Hamiltonian includes an external confining potential $U_{\rm ext}$, such as that induced by metallic gates in electrostatically-defined quantum dots, which adds to the nuclear contribution: $U = U_{\rm n} + U_{\rm ext}$. We remark that, for consistency, the external potential must be expressed in the curvilinear coordinates $\boldsymbol{r}$; i.e., if the potential is initially known as a function $U_{\rm ext; \,C}(\boldsymbol{r}_{\rm C})$ of the Cartesian electronic coordinates, then the expression to be used here is $U_{\rm ext}(\boldsymbol{r}) = U_{\rm ext; \,C}[\boldsymbol{r} + \boldsymbol{u}(\boldsymbol{r})]$. Therefore, this quantity depends on the strain tensor through $\boldsymbol{u}(\boldsymbol{r})$. The external potential is assumed to be a slowly-varying function of $\boldsymbol{r}$ on the scale of the lattice constant, so as to justify an envelope-function approach. 
In particular, the eigenfunctions of $\mathcal{H}$ can be written as:
\begin{align}
    \phi(\boldsymbol{r}) \equiv \big< \boldsymbol{r} \big| \phi \big> =  \frac{1}{\left[-g(\boldsymbol{r}) \right]^{1/4}} \sum_{n \leq N}  F_n(\boldsymbol{r}) \, u_{n, \boldsymbol{0}}(\boldsymbol{r}) \,,
    \label{wfc curvilinear coordinates}
\end{align}
where the $N$ quantities denoted as $F_n(\boldsymbol{r})$ are the unknown envelope functions. These are determined by diagonalizing in band and position spaces the envelope-function Hamiltonian $\hat{\boldsymbol{\mathcal{H}}}_{\rm EF} = \hat{\boldsymbol{\mathcal{H}}}_{\rm EF}^{(0)} + \hat{\boldsymbol{\mathcal{H}}}_{\rm EF}^{(1)}$, where $\hat{\boldsymbol{\mathcal{H}}}_{\rm EF}^{(0)}$ is formally the standard $\boldsymbol{k} \cdot \boldsymbol{p}$ term in the $\boldsymbol{r}$ coordinates, and $\hat{\boldsymbol{\mathcal{H}}}_{\rm EF}^{(1)}$ is given by: 
\begin{widetext}
\begin{align}
    \hat{\boldsymbol{\mathcal H}}_{\rm EF}^{(1)}   =    \varepsilon^{\, \mu}_{\nu}(\boldsymbol{r})   \left[  \boldsymbol{\mathcal{D}}_{\mu}^{\nu}        
    +    \left(  \boldsymbol{\mathcal{L}}_{\alpha; \mu}^{\nu}      +  \boldsymbol{\mathcal{L}}_{\alpha; \mu}^{*\nu} \right) \hat{k}_{\alpha}       +   \boldsymbol{\mathcal{Q}}_{\alpha \beta; \mu}^{\nu}     \hat{k}_{\alpha} \hat{k}_{\beta}        \right]
  -  {\rm i} \left[ \partial_\alpha\varepsilon^{\, \mu}_{\nu}(\boldsymbol{r}) \right]   \left(    \boldsymbol{\mathcal{L}}_{\alpha; \mu}^{\nu}       +  \boldsymbol{\mathcal{Q}}_{\alpha \beta; \mu}^{\nu}     \hat{k}_{\beta}      \right)      + \frac{\hbar^2}{4 m} \left[ \nabla^2 \varepsilon^{\mu}_{\mu}(\boldsymbol{r}) \right]\, \boldsymbol{1}     \,,
   \label{secular equation envelope functions final}
\end{align} 
\end{widetext}
with $\hat{k}_{\alpha} \equiv - {\rm i} \partial_\alpha$. Here $\boldsymbol{\mathcal{D}}$, $\boldsymbol{\mathcal{L}}$, and $\boldsymbol{\mathcal{Q}}$ are matrices in band space, whose elements are defined according to Eqs.~\eqref{matrix element nn'} and \eqref{total deformation potentials}-\eqref{relativistic deformation potentials}; $\boldsymbol{1}$ is the identity matrix in band space.

It should be emphasized that Eq.~\eqref{wfc curvilinear coordinates} is the spinor wave function in the curvilinear reference frame $\boldsymbol{r}$, while the wave function in the Cartesian frame is given by $\phi_{\rm C}(\boldsymbol{r}_{\rm C}) = \phi [\boldsymbol{r}(\boldsymbol{r}_{\rm C})] $, where $\boldsymbol{r}(\boldsymbol{r}_{\rm C})$ is the inverse of the transformation Eq.~\eqref{Jacobian}. Further details on the derivation of the envelope-function Hamiltonian are given in Appendix \ref{app: envelope}.

Equation \eqref{secular equation envelope functions final} is the main result of this work. It contains terms that have not been considered in the literature, and that cannot be inferred from the homogeneous case by replacing a constant strain tensor with a position-dependent one. These terms can be either intra- ($n=n'$) or inter-band ($n \neq n'$), and depend on the first or on the second spatial derivatives of the strain tensor. Besides, the terms linear in $\varepsilon(\boldsymbol{r})$ and $\propto \hat{k}_{\alpha}$ or $\propto \hat{k}_{\alpha} \hat{k}_{\beta}$ are non-zero also in the case of homogeneous strain, but have been neglected in previous analyses.

\section{Valence states in diamond structures} 

As a specific but practically relevant application, we consider the valence bands of a crystal with diamond structure, such as silicon or germanium. In this case, the three relevant orbital states are built from $p$-type atomic orbitals, and thus carry an angular momentum $l=1$. This, combined with the $s=1/2$ spin of the electron, gives rise to a $j=3/2$ quartet and a $j=1/2$ doublet ($N=6$). In this basis, the dominant part of the spin-orbit is diagonal, and gives rise to a splitting $\Delta_{\rm SO}$ between the $j = 3/2$ and $j = 1/2$ states at the $\boldsymbol{\Gamma}$ point \cite{YuCardona}. We here discuss the Hamiltonian that is obtained after neglecting the other spin-orbit contributions. Then: $\mathcal{L}$ coincides with $L$, which vanishes at the band maximum; $\mathcal{D}$ reduces to the non-relativistic deformation potentials $D$; $\mathcal{Q}$ equals $Q$, which consists of purely intraband ($n=n'$) contributions. The strain-dependent component of the envelope-function Hamiltonian matrix, Eq.~\eqref{secular equation envelope functions final}, thus becomes:
\begin{align}
\hat{\boldsymbol{\mathcal{H}}}_{\rm EF}^{(1)} &  \approx  \left\{    \frac{\hbar^2}{4 m} \left[ \nabla^2 \varepsilon^{\mu}_{\mu}(\boldsymbol{r}) \right]  - \frac{\hbar^2  }{m}   
     \varepsilon^{\rm sym }_{\alpha \beta}(\boldsymbol{r})       \hat{k}_{\alpha} \hat{k}_{\beta}      \right. \nonumber\\ & \quad - \left.\frac{\hbar^2  }{m} \left[ \hat{k}_{\alpha}  \varepsilon^{\rm sym}_{\alpha \beta}(\boldsymbol{r}) \right]  \hat{k}_{\beta} 
    \right\}     \mathbf{1}  +        \varepsilon^{\, \mu}_{\nu}(\boldsymbol{r}) \,  \boldsymbol{D}_{\mu}^{\nu}    \,.
\label{6x6 matrix}
\end{align}
Here, $\varepsilon^{\rm sym}_{\alpha \beta}\equiv\frac{1}{2}(\varepsilon^{\alpha}_{\beta}+\varepsilon^{ \beta}_{\alpha})$ is the symmetric part of the strain tensor, and $\boldsymbol{D}_{\mu}^{\nu}$ is the matrix of non-relativistic deformation potentials.

The three main new terms in Eq.~\eqref{6x6 matrix} are diagonal in the band index. The first one, $\propto \nabla^2 \varepsilon^{\mu}_{\mu}(\boldsymbol{r})$, is a function of coordinates only, hence it represents an effective correction to the electrostatic potential, which might affect the hole confinement in a nanostructure. Now, the trace of the strain tensor is proportional to that of the stress tensor, and it can be shown \cite{TimoshenkoGoodier} that the Laplacian of the latter vanishes in an \emph{isotropic} system. However, crystals are not isotropic, and the deviation from isotropy is responsible for the existence of this term. The second term, $\propto \hat{k}_{\alpha} \hat{k}_{\beta}$, represents a spatially-dependent correction to the effective-mass terms, where the spatial dependence is due to the inhomogeneity of the strain tensor. This term is also nonzero in the case of an homogeneous strain; however, it has been ignored so far. The third term, $\propto \hat{k}_{\beta}$, has a completely new form, being linear in the momentum operator. It is not analogous to a spin-orbit coupling, since it does not couple different bands.

In order to provide an order-of-magnitude estimate of the new terms, we consider the stress tensor used in Ref.~\cite{Bonera13}, adapted to describe a MOSFET with two stressors placed on top of the source and drain regions \cite{Bellentani21a} (details are given in Appendix \ref{app: numerics}). We find that, in such a system, the typical energy scales characterizing the new terms is $10^{-2}$ meV, corresponding to frequencies of the order of $2 \div 5$ GHz. In particular, the terms quadratic in the momentum 
induce a spatial modulation of the hole effective masses in Si of approximately $1 \% \div 2 \%$. The hard-to-control modulation of the confinement induced by the term $\propto \nabla^2 \varepsilon^{\mu}_{\mu}(\boldsymbol{r})$ is small and mostly relevant far from the center of the channel (close to the stressors); therefore, it likely has a small impact on the effective confinement. However, the impact of all the derived terms is ultimately device- and material- dependent: they should thus be accounted for in order to obtain an accurate modeling of all nanostructures where the strain tensor varies on the length scale of the particle wave function.

\section{Conclusions} 

By combining solid-state theory and relativistic quantum mechanics in a non-Cartesian geometry, we have derived the envelope-function Hamiltonian for a general semiconductor nanostructure subjected to a small and slowly-varying inhomogeneous strain. Our theory requires, as an input, the strain tensor, which can be computed for each given device via finite-element methods based on the minimization of the elastic energy density \cite{Abadillo23a}. Numerical calculations of the electron/hole states based on our theory are expected to provide an accurate modelling of the effects of inhomogeneous strain on quantum-dot spin qubits. In particular, they will allow to engineer spin-orbit interactions and $g$-tensor modulations aimed at improving the qubits' manipulability.

\acknowledgments

The authors acknowledge financial support from the European Commission through the project IQubits (H2020-FETOPEN-2018-2019-2020-01, Project No.~829005) and from the PNRR MUR Project No.~PE0000023-NQSTI, and useful discussions with S. Pittalis.

\appendix

\section{Inhomogeneous strain and expansion of the nuclei potential}
\label{app: preliminary}

In the literature, there are two definitions of the strain tensor. The first one, which we adopt in the present work, is given by Eq.~\eqref{definition of strain} 
and is consistent with that used in other envelope-function treatments \cite{BirPikus, Zhang94a, VoonWillatzen}. The strain tensor reported in the second definition is the symmetrized version of that given in the first one: 
\begin{align}
    \varepsilon_{\alpha, \beta}^{\rm sym}(\boldsymbol{r}) \equiv \frac{1}{2} \left[ \frac{\partial u^{\alpha}(\boldsymbol{r})}{\partial r^{\beta}} + \frac{\partial u^{\beta}(\boldsymbol{r})}{\partial r^{\alpha}} \right] = \frac{1}{2} \left[ \varepsilon^{\alpha}_{\beta}(\boldsymbol{r}) + \varepsilon^{\beta}_{\alpha}(\boldsymbol{r}) \right] \,.
    \label{symmetric strain}
\end{align}
This is the quantity that enters the expression of the infinitesimal variation in the distance between two points, in going from an unstrained to a strained system \cite{Sverdlov}. The two definitions do not necessarily coincide, since in general $\varepsilon^{\alpha}_{\beta}(\boldsymbol{r}) \neq \varepsilon_{\alpha}^{\beta}(\boldsymbol{r})$. Within the second convention, the quantity introduced in Eq.~\eqref{definition of strain} is termed the {\it deformation tensor}, the symmetric quantity in Eq.~\eqref{symmetric strain} is termed the {\it strain tensor}, while the antisymmetric combination
\begin{align}
    \varepsilon_{\alpha, \beta}^{\rm antisym}(\boldsymbol{r}) \equiv   \frac{1}{2} \left[ \varepsilon^{\alpha}_{\beta}(\boldsymbol{r}) - \varepsilon^{\beta}_{\alpha}(\boldsymbol{r}) \right] \,.
    \label{antisymmetric strain}
\end{align}
is called the {\it rotation tensor}.

The formal solution of Eq.~(2),
as noticed also in Ref.~\cite{Zhang94a}, is
\begin{align}
    u^{\alpha}(\boldsymbol{r}) = u^{\alpha}(\boldsymbol{r}_0) + \int_{\boldsymbol{r}_0}^{\boldsymbol{r}} \varepsilon^{\alpha}_{\beta}(\boldsymbol{r}')   d r'^{\beta} \,,
    \label{relation displacement and strain}
\end{align}
where the line integral is performed over any path connecting $\boldsymbol{r}_0$ to $\boldsymbol{r}$. In the specific case of an homogeneous strain, considered by Bir and Pikus \cite{BirPikus}, this reduces to
\begin{align}
    u^{\alpha}(\boldsymbol{r}) = \varepsilon^{\alpha}_{\beta}   r^{\beta} \,.
\end{align}

In the absence of strain, the potential generated by the nuclei is periodic, and is given by
\begin{align}
    U_{{\rm n}, 0}(\boldsymbol{r}_{\rm C})    = \sum_i U_{1\rm n}\left(\boldsymbol{r}_{\rm C}  -   \boldsymbol{R}_{i, 0}  \right)  \,.
\end{align} 
The same quantity can be rewritten as
\begin{align}
    U_{{\rm n}, 0}(\boldsymbol{r}_{\rm C})     = \sum_{i} \overline{U}_{1 \rm n}\left(\boldsymbol{r}_{\rm C}  -   \boldsymbol{R}_{i, 0} \right) \, \Theta\!\left[ \boldsymbol{r}_{\rm C} \in \mathcal{C}_{0}( \boldsymbol{R}_{i, 0}  ) \right]   \,,
\end{align} 
where $\overline{U}_{1 \rm n}$ is a pseudopotential, and the contribution due to the nucleus at $\boldsymbol{R}_{i, 0}$ goes to zero outside the unit cell $\mathcal{C}_{0}( \boldsymbol{R}_{i, 0}  )$, centered on the same nucleus. This amounts to a mere re-summation of contributions due to all nuclei, and can be done in the strained system as well. In the rigid-ion approximation, with reference to the strained unit cells $\mathcal{C}\left[ \boldsymbol{R}_{i, 0} +  \boldsymbol{u}( \boldsymbol{R}_{i, 0} ) \right]$, the nuclei potential is written as
\begin{align}
    U_{{\rm n}; \, {\rm C}}(\boldsymbol{r}_{\rm C})     = \sum_{i} & \,\, \overline{U}_{1 \rm n}\left[\boldsymbol{r}_{\rm C}  -   \boldsymbol{R}_{i, 0} -  \boldsymbol{u}( \boldsymbol{R}_{i, 0} ) \right] \nonumber \\
    & \times \Theta \! \left\{ \boldsymbol{r}_{\rm C} \in \mathcal{C}\left[  \boldsymbol{R}_{i, 0} +  \boldsymbol{u}( \boldsymbol{R}_{i, 0} ) \right] \right\} .
\end{align} 
After the coordinate transformation $\boldsymbol{r}_{\rm C} = \boldsymbol{r} + \boldsymbol{u}(\boldsymbol{r}) \equiv \boldsymbol{t}(\boldsymbol{r})$, setting $U_{{\rm n}; \, {\rm C}}(\boldsymbol{r}_{\rm C}) = U_{{\rm n}; \, {\rm C}}[\boldsymbol{t}(\boldsymbol{r})] \equiv U_{{\rm n}}(\boldsymbol{r})$, one obtains
\begin{align}
    U_{{\rm n}}(\boldsymbol{r})   
        = \sum_{ i }  & \,\,  \overline{U}_{1 \rm n}\left[   \boldsymbol{r} +  \boldsymbol{u}(\boldsymbol{r})    -   \boldsymbol{R}_{i, 0} -  \boldsymbol{u}( \boldsymbol{R}_{i, 0} )    \right]   \nonumber \\
      & \times \Theta \! \left\{ \boldsymbol{t}(\boldsymbol{r})  \in \mathcal{C}\left[ \boldsymbol{t}(\boldsymbol{R}_{i, 0}) \right] \right\} \,.
\end{align} 
Since the strain, besides being small, varies slowly within a unit cell, the condition imposed by the function $\Theta$ is satisfied only when $\boldsymbol{r}$ is close to $\boldsymbol{R}_{i, 0}$ and, therefore, $\boldsymbol{u}(\boldsymbol{r})      -  \boldsymbol{u}( \boldsymbol{R}_{i, 0} )$ is small. Therefore, the following approximation holds:
\begin{align}
    u^{\nu}(\boldsymbol{r})      -  u^{\nu}( \boldsymbol{R}_{i, 0} ) & \approx \frac{\partial u^{\nu}( \boldsymbol{r} )}{\partial r^{\mu} } \left( r^{\mu} - R^{ \mu}_{i, 0} \right)  \nonumber \\
    & \equiv   \varepsilon^{\nu}_{\mu}\left( \boldsymbol{r} \right) \left( r^{\mu} - R^{ \mu}_{i, 0} \right) \,.
\end{align}
From this it follows that
\begin{align}
     &  \overline{U}_{1 \rm n}\left[   \boldsymbol{r} +  \boldsymbol{u}(\boldsymbol{r})    -   \boldsymbol{R}_{i, 0} -  \boldsymbol{u}( \boldsymbol{R}_{i, 0} )    \right]   \nonumber \\
     &   \approx \overline{U}_{1 \rm n}\left\{ \left[ \delta^{\alpha}_{\beta}  + \varepsilon^{\alpha}_{\beta}\left( \boldsymbol{r}  \right) \right]   \left( r^{\beta}   -   R^{\beta}_{i, 0} \right) \right\}    \nonumber \\
    & \approx   \overline{U}_{1 \rm n}\left(   \boldsymbol{r}   -   \boldsymbol{R}_{i, 0}    \right)      +  \frac{\partial \overline{U}_{1 \rm n}\left( \boldsymbol{r} - \boldsymbol{R}_{i, 0} \right)}{\partial \left( r^{\alpha} - R^{\alpha}_{i, 0} \right)} \, \varepsilon^{\alpha}_{\beta}\left( \boldsymbol{r}  \right) 
 \, \left( r^{\beta} - R^{\beta}_{i, 0} \right)     .
\end{align}
Summing over the atoms, one obtains the total potential as in Eq.~\eqref{expansion of EN potential}, where
\begin{align}
    U^{\beta}_{\alpha}(\boldsymbol{r} ) \equiv \sum_{i} \frac{\partial \overline{U}_{1 \rm n}\left( \boldsymbol{r} - \boldsymbol{R}_{i, 0} \right)}{\partial \left( r^{\alpha} - R^{\alpha}_{i, 0} \right)}   \, \left( r^{\beta} - R^{\beta}_{i, 0} \right)   
  \label{V beta alpha}
\end{align}
has the same periodicity as the unstrained lattice.

\section{Concepts related to point transformations}
\label{app: point transf}

The Schr\"odinger equation,
\begin{align}
    \hat{H} \big| \psi \big> = E \big| \psi \big> \,,
    \label{Schroedinger}
\end{align}
can be written in position representation after introducing a spatial coordinate frame. The purpose of this Appendix is to compare the representations corresponding to two different spatial coordinate frames $\mathbb{R}$ and $\mathbb{R}'$, connected by a spatial transformation. In the first (unprimed) frame, positions are measured by the coordinates $\boldsymbol{r}$, and a position eigenstate is defined by
\begin{align}
    \hat{\boldsymbol{r}} \big| \mathbb{R} ; \boldsymbol{r} \big > = \boldsymbol{r} \big| \mathbb{R} ; \boldsymbol{r} \big >  \,,  
\end{align}
where $\hat{\boldsymbol{r}}$ is the position operator in $\mathbb{R}$. The decomposition of the identity in coordinate space is \cite{DeWitt52a}
\begin{align}
    \int d \Omega \big| \mathbb{R} ; \boldsymbol{r} \big> \big< \mathbb{R} ; \boldsymbol{r} \big| = \hat{1} \,, 
\end{align}
where $d \Omega \equiv d\boldsymbol{r} \sqrt{-g(\boldsymbol{r})}$ is the elementary volume in coordinate space, $g(\boldsymbol{r})$ being the determinant of the metric tensor. The orthonormality relation between the position eigenstates in the $\mathbb{R}$ reference frame is
\begin{align}
    \big< \mathbb{R} ; \boldsymbol{r}_1 \big| \mathbb{R} ; \boldsymbol{r}_2 \big>   \equiv \delta (\boldsymbol{r}_1, \boldsymbol{r}_2)   \equiv \frac{1}{\sqrt{- g(\boldsymbol{r}_1)}} \,\delta(\boldsymbol{r}_1 - \boldsymbol{r}_2)   \,.
\end{align}
Multiplying Eq.~\eqref{Schroedinger} by $\big< \mathbb{R} ; \boldsymbol{r} \big|$, one obtains the usual coordinate-space Schr\"odinger problem:
\begin{align}
      H(\boldsymbol{r}) \,\psi(\boldsymbol{r}) = E\, \psi(\boldsymbol{r}) \,,
    \label{Schroedinger original coordinates}
\end{align}
where
\begin{align}
    & \psi(\boldsymbol{r}) \equiv \big< \mathbb{R} ; \boldsymbol{r} \big| \psi \big>  \,, \quad \big< \mathbb{R} ; \boldsymbol{r} \big| \hat{H} \big| \psi \big> \equiv H(\boldsymbol{r}) \,\psi(\boldsymbol{r})  \,, \nonumber \\
    & \big< \mathbb{R} ; \boldsymbol{r}_1 \big| \hat{H} \big| \mathbb{R} ; \boldsymbol{r}_2 \big> \equiv \delta(\boldsymbol{r}_1, \boldsymbol{r}_2)\, H(\boldsymbol{r}_2) \,,
\end{align}
and the eigenstate of the Hamiltonian in Hilbert space representation is
\begin{align}
    \big| \psi \big> = \int d \Omega\,\big| \mathbb{R} ; \boldsymbol{r} \big>\, \psi(\boldsymbol{r}) \,.
\end{align}

To solve the same Schr\"odinger problem as in Eq.~\eqref{Schroedinger}, one can equivalently choose the different coordinate frame $\mathbb{R}'$, with coordinates $\boldsymbol{r}'$, connected to the previous representation via $\boldsymbol{r} = \boldsymbol{t}(\boldsymbol{r}')$. It should be noted that this is a relation between the position {\it eigenvalues} measured in different reference frames on the same position eigenstate. This means that the following relations between the position eigenstates in the two reference frames hold:
\begin{align}
    & \big| \mathbb{R} ; \boldsymbol{t}(\boldsymbol{r}') \big> = \big| \mathbb{R}' ; \boldsymbol{r}' \big>  \,, \quad
    \big| \mathbb{R} ; \boldsymbol{r} \big> = \big| \mathbb{R}' ; \boldsymbol{t}^{-1}(\boldsymbol{r})  \big>  \,.
    \label{correspondence between reference frame position eigenstates}
\end{align}
In other words, if a position measurement on a position eigenstate gives the result $\boldsymbol{r}'$ in the reference frame $\mathbb{R}'$, then the result of the position measurement on the same state gives the result $\boldsymbol{t}(\boldsymbol{r}')$ in the reference frame $\mathbb{R}$:
\begin{align}
    & \hat{\boldsymbol{r}} \,\big| \mathbb{R} ; \boldsymbol{t}(\boldsymbol{r}') \big> = \boldsymbol{t}(\boldsymbol{r}')\, \big| \mathbb{R} ; \boldsymbol{t}(\boldsymbol{r}') \big> \,, \quad \!
    \hat{\boldsymbol{r}}' \,\big| \mathbb{R}' ; \boldsymbol{r}' \big> = \boldsymbol{r}'\, \big| \mathbb{R}' ; \boldsymbol{r}' \big>  \,,
\end{align}
where $\hat{\boldsymbol{r}}'$ is the position operator in $\mathbb{R}'$.  

In the primed representation, the identity decomposition is
\begin{align}
    \int d \Omega'\,\big| \mathbb{R}' ; \boldsymbol{r}' \big> \big< \mathbb{R}' ; \boldsymbol{r}' \big| = \hat{1} \,, \quad \big< \mathbb{R}' ; \boldsymbol{r}'_1 \big| \mathbb{R}' ; \boldsymbol{r}'_2 \big> = \delta'(\boldsymbol{r}'_1, \boldsymbol{r}'_2) \,.
\end{align}
The Schr\"odinger problem is represented in $\mathbb{R}'$ as
\begin{align}
      H'(\boldsymbol{r}')\, \psi'(\boldsymbol{r}') = E\, \psi'(\boldsymbol{r}') \,,
    \label{Schroedinger primed coordinates}
\end{align}
with 
\begin{align}
    \big| \psi \big> = \int d \Omega' \big| \mathbb{R}' ; \boldsymbol{r}' \big> \psi'(\boldsymbol{r}') \,.
\end{align}

One should now derive the relation between $\psi(\boldsymbol{r})$ and $\psi'(\boldsymbol{r}')$, and that between $H(\boldsymbol{r})$ and $H'(\boldsymbol{r}')$.
 The most direct way to do so is to consider the scalar product between any two states $\big| \Phi_1 \big>$ and $\big| \Phi_2 \big>$, which must be the same independently of the reference frame where it is evaluated:
\begin{align}
    \big< \Phi_1 \big| \Phi_2 \big>   = \! \int \! d \Omega \, [\Phi_1(\boldsymbol{r})]^* \, \Phi_2(\boldsymbol{r})   = \! \int \! d \Omega' \, [\Phi_1'(\boldsymbol{r}')]^* \, \Phi'_2(\boldsymbol{r}')\,.
\end{align}
We now change variables in the first equality according to $\boldsymbol{r} = \boldsymbol{t}(\boldsymbol{r}')$, using the transformation property of the determinant of the metric tensor,
\begin{align}
    \sqrt{- g'(\boldsymbol{r}')} = \left| J(\boldsymbol{r}') \right| \sqrt{-g[\boldsymbol{t}(\boldsymbol{r}')]} \,,
\end{align}
where $J(\boldsymbol{r}')$ is the determinant of the Jacobian matrix. Therefore, under this change of coordinates it holds that
\begin{align}
    \int d \Omega \, \{\Phi_1(\boldsymbol{r})\}^* \, \Phi_2(\boldsymbol{r})   &  = \int d \Omega' \, \{\Phi_1[\boldsymbol{t}(\boldsymbol{r}')]\}^* \, \Phi_2[\boldsymbol{t}(\boldsymbol{r}')] \nonumber \\
    & = \int d \Omega' \, [\Phi_1'(\boldsymbol{r}')]^* \, \Phi'_2(\boldsymbol{r}') \,.
\end{align}
Since this must hold for any couple of states $\big| \Phi_1 \big>$ and $\big| \Phi_2 \big>$, one concludes that
\begin{align}
    \Phi'(\boldsymbol{r}') = \Phi[\boldsymbol{t}(\boldsymbol{r}')] \quad {\rm if} \,\, \boldsymbol{r} = \boldsymbol{t}(\boldsymbol{r}') \,.
    \label{transformation of wave functions}
\end{align}
Therefore,
\begin{align}
    \psi'(\boldsymbol{r}') = \psi[\boldsymbol{t}(\boldsymbol{r}')] \quad {\rm if} \,\, \boldsymbol{r} = \boldsymbol{t}(\boldsymbol{r}') \,,
\end{align}
which provides the relation between the wave functions in the two reference frames.

In order to derive an analogous relation between the Hamiltonians, let us consider again Eq.~\eqref{correspondence between reference frame position eigenstates}. Multiplying both sides of the first equation by $\big< \mathbb{R}; \boldsymbol{r} \big|$, and using the orthonormality of the position eigenstates in $\mathbb{R}$, one obtains
\begin{align}
      \big< \mathbb{R}; \boldsymbol{r} \big| \mathbb{R}' ; \boldsymbol{r}' \big> = \delta\left[\boldsymbol{r}, \boldsymbol{t}(\boldsymbol{r}')\right] = 
      \frac{\delta[\boldsymbol{r} - \boldsymbol{t}(\boldsymbol{r}')]}{\sqrt{-g(\boldsymbol{r})}} \,.
    \label{<x|x'>}
\end{align}
Analogously, multiplying both sides of the second equation by $\big< \mathbb{R}'; \boldsymbol{r}' \big|$, and using the orthonormality of the position eigenstates in $\mathbb{R}'$, one obtains
\begin{align} 
    & \big< \mathbb{R}'; \boldsymbol{r}' \big| \mathbb{R} ; \boldsymbol{r} \big> = \delta'\left[ \boldsymbol{r}' , \boldsymbol{t}^{-1}(\boldsymbol{r})  \right] = \frac{\delta\left[ \boldsymbol{r}' -  \boldsymbol{t}^{-1}(\boldsymbol{r})  \right]}{\sqrt{- g'(\boldsymbol{r}')}} \,.
    \label{<x'|x>}
\end{align}
The two transformation laws between position eigenstates in the two reference frames are equivalent, due to the composition law of a Dirac delta with a function. Equipped with Eqs.~\eqref{<x|x'>} and \eqref{<x'|x>}, it is now possible to compare the quantities
\begin{align}
    H(\boldsymbol{r}_1 , \boldsymbol{r}_2 ) \equiv \big< \mathbb{R} ; \boldsymbol{r}_1 \big| \hat{H} \big| \mathbb{R} ; \boldsymbol{r}_2 \big> 
    \label{H x1 x2}
\end{align}
and 
\begin{align}
    H'(\boldsymbol{r}'_1 , \boldsymbol{r}'_2 ) \equiv \big< \mathbb{R}' ; \boldsymbol{r}'_1 \big| \hat{H} \big| \mathbb{R}' ; \boldsymbol{r}'_2 \big> \,. 
    \label{H' x'1 x'2}
\end{align}
Inserting twice the decomposition of the identity in the original reference frame in Eq.~\eqref{H' x'1 x'2} and using the scalar product relations derived above, one obtains
\begin{align}
    & H'(\boldsymbol{r}'_1 , \boldsymbol{r}'_2 ) \nonumber \\
    & = \int d \Omega_1 \int d \Omega_2\, \big< \mathbb{R}' ; \boldsymbol{r}'_1 \big| \mathbb{R} ; \boldsymbol{r}_1 \big>\, H(\boldsymbol{r}_1, \boldsymbol{r}_2) \,\big< \mathbb{R} ; \boldsymbol{r}_2  \big| \mathbb{R}' ; \boldsymbol{r}'_2 \big>  \nonumber \\
    & = H[\boldsymbol{t}(\boldsymbol{r}'_1), \boldsymbol{t}(\boldsymbol{r}'_2)] \,,
    \label{transformation Hamiltonian}
\end{align}
which provides the relation between the Hamiltonians in the Cartesian and in the curvilinear coordinates.

\section{Derivation of the covariant Schr\"odinger equation, with relativistic corrections}
\label{app: derivation Schr}

In this Appendix, Greek indexes refer to the components of vectors and tensors in a curvilinear reference frame, while Latin indexes refer to such components in a Cartesian (rectilinear) reference frame, which coincides with the laboratory reference frame of the main text (note that in the main text Greek indexes are used for both frames). The 4-component space-time coordinate is denoted as $x$ in the curvilinear frame, and as $x_{\rm C}$ in the Cartesian frame. The metric tensor in the $x$ frame is denoted as $g_{\mu \nu}$, while the metric tensor in the $x_{\rm C}$ frame is $\eta_{ab} = {\rm diag}(1, -1, -1, -1)$. The relation between the two metric tensors is \cite{Breev16a}
\begin{align}
    g_{\mu \nu} = e^a_{\mu}\, e^b_{\nu}\, \eta_{ab} \,,  
\end{align}
where the tetrad fields $e^{\mu}_a$ satisfy 
\begin{align}
    e^{\mu}_a e^a_{\nu} = \delta^{\mu}_{\nu} \,, \quad e^{\mu}_a e^b_{\mu} = \delta^{b}_{a} \,.
\end{align}
In the case at hand, since the two coordinate frames are connected by a point transformation, one has
\begin{align}
    e^{a}_{\mu} = \frac{\partial x^a_{\rm C}}{\partial x^{\mu}} \,, \quad  e_{a}^{\mu} = \frac{\partial x^{\mu}}{\partial x^a_{\rm C} } \,, 
    \label{tetrads as Jacobian}
\end{align}
i.e., the tetrad vectors coincide with the Jacobian matrix of the transformation between the two coordinate systems. This ensures the conservation of the infinitesimal arc length squared,
\begin{align}
    \eta_{ab}\, dx^a_{\rm C}\, dx^b_{\rm C} = g_{\mu \nu}\, dx^{\mu}\, dx^{\nu} \,,
\end{align}
which must be independent of the chosen coordinates.

\subsection{Covariant Dirac equation}

We start from the covariant Dirac equation \cite{Breev16a} for the 4-component electron field $\Psi$,
\begin{align}
     \gamma^{\mu} \left( {\rm i} \hbar \nabla_{\mu} - e A_{\mu} \right) \Psi - m c \Psi = 0 \,,
\end{align}
where $A_{\mu}$ is the 4-potential of the electromagnetic field, and $\nabla_{\mu}$ is the covariant derivative; the latter acts on the 4-spinor as follows:
\begin{align}
    \nabla_{\mu} \Psi   \equiv  \partial_{\mu} \Psi + \Gamma_{\mu} \Psi   \,,
\end{align}
where $\partial_{\mu}$ is the ordinary derivative, and
\begin{align}
    & \Gamma_{\mu}   \equiv \frac{1}{2} \eta_{a c} e^{c}_{\nu} \left( \partial_{\mu} e^{\nu}_b + e^{\rho}_b \Gamma^{\nu}_{\rho \mu} \right) G^{ab}  \,,   \nonumber \\
    & G^{ab} = \frac{1}{4} \left( \gamma^a \gamma^b - \gamma^b \gamma^a \right) \,.
\end{align}
The coordinate-invariant Dirac matrices satisfy
\begin{align}
    \gamma^a \gamma^b + \gamma^b \gamma^a = 2 \eta^{ab} \,,
\end{align}
while the spacetime-dependent Dirac matrices (entering the covariant Dirac equation) are defined as $\gamma^{\mu} = e^{\mu}_a \gamma^a$; thus, they satisfy
\begin{align}
    \gamma^{\mu} \gamma^{\nu} + \gamma^{\nu} \gamma^{\mu} = 2 g^{\mu \nu} \,.
\end{align}

In terms of the tetrads, the Christoffel symbols are written as:
\begin{widetext}
\begin{align}
    \Gamma^{\nu}_{\rho \mu}  \equiv \frac{1}{2} g^{\nu \beta} \left(  \partial_{\mu} g_{\beta \rho}   +   \partial_{\rho} g_{\beta \mu}  -   \partial_{\beta} g_{\rho \mu}   \right)  
    = \frac{1}{2}      e^{\nu}_i     \left(  \partial_{\mu}  e^i_{\rho}     +    \partial_{\rho}  e^i_{\mu}    \right)     + \frac{1}{2}  \eta^{kl} \eta_{ij}   e^{\nu}_k e^{\beta}_l \left[ 
     e^i_{\rho} \left(  \partial_{\mu} e^j_{\beta}     -     \partial_{\beta}  e^j_{\mu}    \right)      +  e^i_{\mu} \left(  \partial_{\rho} e^j_{\beta}     
    -    \partial_{\beta} e^j_{\rho}      \right) 
     \right]  \,.
     \label{Christoffel in terms of tetrads - general}
\end{align}    
\end{widetext}
In the case at hand, since Eq.~\eqref{tetrads as Jacobian} holds, one has
\begin{align}
    \partial_{\mu} e^i_{\rho} = \frac{\partial^2 x_{\rm C}^i}{\partial x^{\mu} \partial x^{\rho}} = \partial_{\rho} e^i_{\mu} \,,
\end{align}
and Eq.~\eqref{Christoffel in terms of tetrads - general} simplifies as
\begin{align}
    \Gamma^{\nu}_{\rho \mu}   =      e^{\nu}_i     \left(       \partial_{\rho}  e^i_{\mu}    \right)    \,.
     \label{Christoffel in terms of tetrads - simplified}
\end{align}
The combinations needed for the covariant derivative are 
\begin{align}
   \frac{1}{2} \eta_{ac} e^c_{\nu} e^{\rho}_b \Gamma^{\nu}_{\rho \mu} G^{ab} 
     = \frac{1}{2} \eta_{ac}   e^{\rho}_b       \left(       \partial_{\rho}  e^c_{\mu}    \right)   G^{ab}   \,,
\end{align}
and
\begin{align}
    \gamma^{\mu} \Gamma_{\mu} & =  \frac{1}{2} \eta_{a c}  \left[  e^{c}_{\nu}   \left( \partial_{d} e^{\nu}_b \right)  +   e^{\mu}_d      \left(       \partial_{b}  e^c_{\mu}    \right)    \right] \gamma^d G^{ab}   \nonumber \\
    & =  \frac{1}{2} \eta_{a c}  \left[  e^{c}_{\mu}   \left( \partial_{d} e^{\mu}_b \right)  -   \left(       \partial_{b} e^{\mu}_d    \right)      e^c_{\mu}       \right] \gamma^d G^{ab}    = 0 \,, 
\end{align}
where we have used the fact that $\partial_{\mu} \left( e^c_{\nu} e^{\nu}_b \right) = 0$, and thus $e^c_{\nu} \partial_{\mu} e^{\nu}_b = - e^{\nu}_b \partial_{\mu} e^c_{\nu}$. Therefore, the covariant Dirac equation, when the tetrad coincides with the Jacobian matrix, reduces to
\begin{align}
     {\rm i} \hbar e^{\mu}_n \gamma^{n} \partial_{\mu} \Psi   - e e^{\mu}_n A_{\mu} \gamma^{n}  \Psi   - m c \Psi = 0 \,.
     \label{Dirac tetrads = jacobian}
\end{align}

In the following, we choose the coordinate-invariant Dirac matrices in the Dirac form, 
\begin{align}
    \gamma^0 = \left( \begin{matrix} 1_{2\times 2} & 0_{2\times 2} \\ 0_{2\times 2} & -1_{2\times 2} \end{matrix} \right) \,;
\quad   \gamma^a = \left( \begin{matrix} 0_{2\times 2} & \sigma^a \\ - \sigma^a & 0_{2\times 2} \end{matrix} \right) \,,  
\end{align}
where  $\sigma^a$, with $a \in \lbrace 1, 2, 3 \rbrace$, are the Pauli matrices.

\subsection{Restriction to spatial-only transformation}

In the problem at hand, the coordinate transformation only involves the spatial coordinates ($\mu \in \lbrace 1, 2, 3 \rbrace$), affecting the corresponding sector of the metric tensor, while the time coordinate ($\mu = 0$) is untouched. Therefore, we specialize our treatment to the cases where $e^0_{\mu} = \delta^0_{\mu}$ and $e^a_{0} = \delta^a_0$. All tetrads are independent of time.   

It is then convenient to write the Dirac equation in a way that explicitly separates time from the spatial coordinates. Equation~\eqref{Dirac tetrads = jacobian} can thus be transformed into 
\begin{align}
       \left( \begin{matrix}  1 & 0 \\ 0 & -1 \end{matrix}\right) {\rm i} \hbar  \partial_t \Psi
     & =        \left( \begin{matrix}
         0 & c \sigma^{\mu} P_{\mu} \\ - c \sigma^{\mu} P_{\mu} & 0
     \end{matrix}\right)   \Psi \nonumber \\
     & \quad 
     +   \left( \begin{matrix}
         m c^2 + e   V & 0 \\ 0 & m c^2 - e   V 
     \end{matrix}\right)  \Psi    \,,
     \label{Dirac covariant space and time separated}
\end{align}
where $x_0 = c t$, $\partial_0 = \frac{1}{c} \partial_t$, $A_0 = \frac{1}{c} V$, $P_{\mu} \equiv -   {\rm i} \hbar   \partial_{\mu}    + e   A_{\mu} $, and the coordinate-dependent Pauli matrices are defined as
\begin{align}
    \sigma^{\mu} \equiv e^{\mu}_n \sigma^n \,.
    \label{coordinate dependent Pauli matrices}
\end{align}

This is rephrased as an eigenproblem, by setting
\begin{align}
    \Psi \equiv {\rm e}^{- {\rm i} E_{\rm D} t / \hbar} \left( \begin{matrix} \Xi \\ \Phi \end{matrix} \right)  ,
\end{align}
where $E_{\rm D}$ is the (Dirac) energy eigenvalue, while $\Xi$ and $\Phi$ are two-component spinors depending only on the spatial coordinates (eigenstates). This results in the following coupled equations for the 2-component spinor eigenstates:
\begin{align}
    & \left( E_{\rm D} - mc^2 - eV \right) \Xi  =  c \sigma^{\mu} P_{\mu}    \Phi   \,, \nonumber \\
    & \left( E_{\rm D} + mc^2 - eV \right) \Phi =  c \sigma^{\mu} P_{\mu}    \Xi    \,.
    \label{coupled equations for spinors}
\end{align}
The spinor $\Phi$ is obtained as a function of $\Xi$ from the second equation; substituting the resulting expression in the first equation, one obtains an eigenvalue equation for the spinor $\Xi$ alone: 
\begin{align}
      \left[     c \sigma^{\mu} P_{\mu}   \frac{1}{2 mc^2 + E   - eV}         c \sigma^{\nu} P_{\nu}   + eV \right] \Xi = E  \, \Xi  \,.
      \label{equation for the relevant spinor}
\end{align}
where the Schr\"odinger eigenenergy is $E \equiv E_{\rm D} - mc^2$. This equation is exact.

\subsection{Schr\"odinger equation with relativistic corrections, in curvilinear coordinates}

In order to recover the Schr\"odinger equation and the lowest-order relativistic corrections, one must perform an approximation based on the assumption that $2 mc^2 \gg |E - eV |$, namely,
\begin{align}
     \frac{1}{2 mc^2 + E   - eV}  & = \frac{1}{2 mc^2} \left( 1 + \frac{E   - eV }{2 mc^2} \right)^{-1} \nonumber \\
     & \approx   \frac{1}{2 mc^2} - \frac{E   - eV }{4 m^2 c^4} \,.
\end{align}
Inserting this into Eq. \eqref{equation for the relevant spinor}, one obtains
\begin{align}
      \left[   \frac{\left(   \sigma^{\mu} P_{\mu} \right)^2 }{2 m }   
      -  \sigma^{\mu} P_{\mu}  \frac{E - eV}{4 m^2 c^2  }      \sigma^{\nu} P_{\nu}  + eV \right] \Xi \approx  E   \, \Xi  \,.
      \label{Schroedinger plus relativistic corrections}
\end{align}
To remove $E$ from the left-hand side of the above equation (to lowest order in $v/c$), we notice that
\begin{align}
    \left( E - eV \right)   \sigma^{\nu} P_{\nu}   \Xi 
    & =  \sigma^{\nu} P_{\nu}  \left( E - eV \right)  \Xi 
    - {\rm i} \hbar \sigma^{\nu} \left(   \partial_{\nu}  eV \right) \Xi \nonumber \\
    & \approx \frac{\left(     \sigma^{\nu} P_{\nu} \right)^3 }{2m}      \Xi 
    - {\rm i} \hbar \sigma^{\nu} \left(   \partial_{\nu}  eV \right) \Xi \,.
\end{align}
Equation \eqref{Schroedinger plus relativistic corrections} can thus be written as an eigenvalue equation
\begin{align}
    H   \, \Xi  \approx E \, \Xi \,,
\end{align}
where the Hamiltonian is given by
\begin{align}
      H & =  \underbrace{ \frac{\left(  \sigma^{\mu} P_{\mu} \right)^2 }{2 m }   + eV  }_{\text{nonrelativistic} }
       \underbrace{  - \frac{\left(   \sigma^{\mu} P_{\mu} \right)^4  }{8 m^3 c^2 }  
      + \frac{{\rm i} \hbar \left(   \sigma^{\mu} P_{\mu} \right)      \sigma^{\nu} \left(   \partial_{\nu}  eV \right)   }{4 m^2 c^2 }    }_{\text{relativistic corrections}}  \nonumber \\
      & \equiv H_{\rm nonrel} + H_{\rm rel}    \,.
      \label{Schroedinger plus relativistic corrections - 2}
\end{align}

\subsection{Hamiltonian in curvilinear coordinates}

Finally, by making the electromagnetic potentials explicit, and using the properties of the Pauli matrices, one can rewrite the terms $H_{\rm nonrel} $ and $ H_{\rm rel}$ appearing in Eq.~\eqref{Schroedinger plus relativistic corrections - 2} as follows:
\begin{align}
      H_{\rm nonrel} & = \frac{1}{2 m}\Big\{  - \hbar^2 \left( - g^{\mu \nu}  \right)  \partial_{\mu}      \partial_{\nu}  - \hbar^2 \left( \nabla^2_{\rm C} x^{\nu} \right)   \partial_{\nu}  \nonumber \\
      & \quad   +  {\rm i} \hbar e  \left[ e^{\mu}_a \left( \partial_{\mu} e^a_{\nu} \right)    A^{\nu}   +    \left( \partial_{\mu} A^{\mu} \right)  + 2  A^{\mu} \partial_{\mu} \right] \nonumber \\
      & \quad -  e^2   A_{\mu}   A^{\mu}     - e \hbar \,   \sigma_c    B^c \Big\}    + eV ,
      \label{Schroedinger curvilinear}\\
      H_{\rm rel} & = - \frac{ 1 }{8 m^3 c^2 } \Big\{ - \hbar^2 \left( - g^{\mu \nu}  \right)  \partial_{\mu}      \partial_{\nu}  - \hbar^2 \left( \nabla^2_{\rm C} x^{\nu} \right)   \partial_{\nu}  \nonumber \\
      & \quad +  {\rm i} \hbar e  \left[ e^{\mu}_a \left( \partial_{\mu} e^a_{\nu} \right)    A^{\nu}   +    \left( \partial_{\mu} A^{\mu} \right)  + 2  A^{\mu} \partial_{\mu} \right] \nonumber \\
      & \quad -  e^2   A_{\mu}   A^{\mu}     - e \hbar \,   \sigma_c    B^c \Big\}^2  \nonumber \\
      & \quad      + \frac{  \hbar^2   }{4 m^2 c^2 }      \left( \nabla^2_{\rm C} eV \right)       - \frac{{\rm i} \hbar   }{4 m^2 c^2 }      \left( \partial_{\mu} eV \right)  g^{\mu \nu}  P_{\nu}  \nonumber \\
      & \quad + \frac{  \hbar   }{4 m^2 c^2 }    \left(    e^{\mu}_a e^{\nu}_b \sigma^{ab} \right) \left( \partial_{\mu} eV \right)  P_{\nu}      \,.
      \label{rel corrections curvilinear}
\end{align} 
In the above expressions, $\sigma_a = \eta_{ab} \sigma^b$, where the implicit summation on the Latin indices only involves the three spatial coordinates. Besides, we have introduced $\sigma^{ab} \equiv - \widetilde{\epsilon}^{\, abc} \sigma_c $ and the components of the magnetic field with respect to the rectilinear reference frame, $B^c \equiv \widetilde{\epsilon}^{\, abc} \left(  \partial_{a}        A_{b} \right)$; here, $\widetilde{\epsilon}^{\, abc}$ is the totally antisymmetric Levi-Civita symbol, which, unlike the Levi-Civita tensor, takes the same values in all reference frames. Finally, we have used the symbol $\nabla^2_{\rm C}$, which can be converted to curvilinear coordinates through:
\begin{align}
    \nabla^2_{\rm C} & = \partial_a \partial_a   = - \partial_a \partial^a = - e^{\alpha}_a \partial_{\alpha} e_{\beta}^a \partial^{\beta} \nonumber \\
    & = - e^{\alpha}_a \left( \partial_{\alpha} e_{\beta}^a \right) \partial^{\beta} -   \partial_{\alpha}  \partial^{\alpha}  \nonumber \\
    & = - e^{\alpha}_a \left( \partial_{\alpha} e_{\beta}^a \right) g^{\beta \gamma} \partial_{\gamma} -   \left( \partial_{\alpha} g^{\alpha \beta} \right) \partial_{\beta}  -    g^{\alpha \beta} \partial_{\alpha} \partial_{\beta}  \,.
\end{align}
Alternatively (and more simply), the relevant derivatives can be computed in the rectilinear coordinate frame first, and then the resulting expressions can be converted in curvilinear coordinates.

In the main text, we consider the case where there is no magnetic field ($A^{\mu} = 0$), and only spin-orbit coupling is retained among the relativistic corrections. In this case, the Hamiltonian simplifies to the sum of the kinetic term given in Eq.~\eqref{Schroedinger curvilinear, no A},
the scalar potential $U \equiv eV$, and the spin-orbit term given in Eq.~\eqref{rel corrections curvilinear, no A}. 
For notational convenience and self-containedness, in the main text we have written the tetrads explicitly in terms of the inverse Jacobian matrix, and we have used Greek indices also for the components of the Cartesian coordinates; since such coordinates themselves are indicated explicitly by means of the subscript C, there is no ambiguity.

\section{Matrix elements of the Hamiltonian in a manifestly Hermitian form}
\label{app: manifestly}

Using the definition of the matrix elements in the presence of an arbitrary metric tensor [Eq.~\eqref{scalar product in arbitrary metric}], one can write the matrix elements of the Hamiltonian between 2-component spinors $\big| \Xi_n \big>$ in a manifestly Hermitian way, assuming that the wave functions either vanish at infinity, or satisfy the Born-von Karman boundary conditions (BvKBCs). The three terms of the Hamiltonian [see Eqs.~\eqref{Schroedinger curvilinear, no A} and \eqref{rel corrections curvilinear, no A}] give the following contributions:
\begin{align}
     & \big< \Xi_n \big| \hat{H}_{\rm kin}  \big| \Xi_m \big>   = - \frac{\hbar^2}{2 m} \int d \boldsymbol{r} \sqrt{-g} g^{\mu \nu} \frac{\partial \Xi^{\dagger}_n}{\partial r^{\mu}} \cdot  \frac{\partial \Xi_m}{\partial r^{\nu}} \,, \label{matrem hermitian1} \\
    & \big< \Xi_n \big|   \hat{U}  \big| \Xi_m \big>   =   \int d \boldsymbol{r} \sqrt{-g} \, U \, \Xi^{\dagger}_n \cdot  \Xi_m  \,, \label{matrem hermitian2} \\
    &  \big< \Xi_n \big| \hat{H}_{\rm so}  \big| \Xi_m \big>   = - \frac{ {\rm i} \hbar^2   }{8 m^2 c^2 }    \int d \boldsymbol{r} \sqrt{- g} \,    \frac{\partial r^{\mu}}{\partial r^{\alpha}_{\rm C} }  \frac{\partial r^{\nu}}{\partial r^{\beta}_{\rm C} }  \left( \partial_{\mu} U \right)  \nonumber \\
    & \quad \quad\quad\quad \quad \times \Big[ \Xi^{\dagger}_n \cdot \sigma^{\alpha \beta}   \cdot   \left( \partial_{\nu} \Xi_m \right)   -  \left( \partial_{\nu} \Xi^{\dagger}_n \right)  \cdot \sigma^{\alpha \beta}   \cdot \Xi_m  \Big] \,.
    \label{matrem hermitian3}
\end{align}
These forms are obtained by applying partial integration, using the boundary conditions, and exploiting the properties of the metric tensor. 

As an example, we show the explicit derivation of the first contribution. One starts from
\begin{align}
    & \big< \Xi_n \big| \hat{H}_{\rm kin}  \big| \Xi_m \big> \nonumber \\
    & = - \frac{\hbar^2}{2 m} \int d \boldsymbol{r} \sqrt{-g} \,\Xi^{\dagger}_n \cdot \left[\left( \nabla_{\rm C}^2 r^{\nu} \right)    - g^{\mu \nu} \frac{\partial}{\partial r^{\mu} }         \right] \frac{ \partial \Xi_m }{\partial r^{\nu}} \,,
\end{align}
which follows from Eq.~\eqref{Schroedinger curvilinear, no A}. Now, partial integration with respect to $r^{\mu}$ is applied to the second term of this integral; the boundary term vanishes due to the boundary conditions, and the remaining term is 
\begin{align}
     \big< \Xi_n \big| \hat{H}_{\rm kin}  \big| \Xi_m \big> 
    & = - \frac{\hbar^2}{2 m} \int d \boldsymbol{r}     \Bigg[ \sqrt{-g} \left( \nabla_{\rm C}^2 r^{\nu} \right) \, \Xi^{\dagger}_n \nonumber \\
    & \quad - \frac{\partial \left(  - g^{\mu \nu} \sqrt{-g} \, \Xi^{\dagger}_n\right) }{\partial r^{\mu} }         \Bigg]    \cdot \frac{ \partial \Xi_m }{\partial r^{\nu}} \,.  
\end{align}
The derivative with respect to $r^{\mu}$ at the second term inside the square brackets is carried on by applying the following identities:
\begin{align}
    & \partial_{\mu}(-g^{\mu \nu}) = - \eta^{ab} e^{c}_{\mu} (\partial_c e^{\mu}_a) e^{\nu}_b + \left( \nabla^2_{\rm C} r^{\nu} \right) \,, \nonumber \\
    &  \partial_{\mu} \sqrt{-g} = \sqrt{-g} e^{\alpha}_a (\partial_{\mu} e^a_{\alpha}) \,, \nonumber \\
    & g^{\mu \nu} e^{\alpha}_a (\partial_{\mu} e^a_{\alpha}) + \eta^{ab} (\partial_c e^{\mu}_a) e^c_{\mu} e^{\nu}_b = 0 \,,
\end{align}
which follow directly from the definitions of the metric tensor and of the tetrads, and from the fact that $\partial_a e^{\alpha}_c = \partial_c e^{\alpha}_a$ in the case at hand, because the tetrads are defined via a point transformation from a Cartesian reference frame. After this, one directly obtains Eq.~\eqref{matrem hermitian1}. Equation~\eqref{matrem hermitian3} follows from a similar derivation.

The symmetry of the matrix elements in Eqs.~(\ref{matrem hermitian1}-\ref{matrem hermitian3}) can be made explicit by formally rewriting the operators themselves as follows:
\begin{align}
    \hat{H}_{\rm kin} = - \frac{\hbar^2}{2 m}  \overleftarrow{\partial}_{\mu} g^{\mu \nu} \overrightarrow{\partial}_{\nu} \,,
    \label{kin hermitian}
\end{align}
\begin{align}
    \hat{H}_{\rm so} & = - \frac{ {\rm i} \hbar^2   }{8 m^2 c^2 }      \Bigg[     \frac{\partial r^{\mu}}{\partial r^{\alpha}_{\rm C} }  \frac{\partial r^{\nu}}{\partial r^{\beta}_{\rm C} }  \left( \partial_{\mu} U \right)       \sigma^{\alpha \beta}        \overrightarrow{\partial}_{\nu}  \nonumber \\
    & \quad - \overleftarrow{\partial}_{\nu}  \frac{\partial r^{\mu}}{\partial r^{\alpha}_{\rm C} }  \frac{\partial r^{\nu}}{\partial r^{\beta}_{\rm C} }  \left( \partial_{\mu} U \right)     \sigma^{\alpha \beta}  \Bigg]  \,,
    \label{so hermitian}
\end{align}
where the arrow above a derivative operator indicates the direction along which the derivative operator acts, when evaluating a matrix element; it is intended that, within this convention, the derivatives do {\it not} act on the metric factor $\sqrt{-g}$ inside the integrals. These two definitions are equivalent to Eqs.~\eqref{Schroedinger curvilinear, no A} and \eqref{rel corrections curvilinear, no A}; the same convention is used in Eqs.~\eqref{Schroedinger Hamiltonian kin} and \eqref{Schroedinger Hamiltonian so}, which are the expansions of Eqs.~\eqref{kin hermitian} and \eqref{so hermitian}, respectively, to the first order in the strain tensor.

\section{Secular equation up to first order in the strain tensor}
\label{app: secular}

We now derive the expressions of the matrix elements of the Hamiltonian up to first order in the strain components on the $\overline{\chi}$ basis, i.e. we derive Eq.~\eqref{matrix element LK}
by evaluating the matrix elements of Eqs.~\eqref{Schroedinger Hamiltonian} and \eqref{strain correction Hamiltonian}. By construction, the $\overline{\chi}$ are orthonormal, i.e., they satisfy
\begin{align}
    \big< \overline{\chi}_{n, \boldsymbol{k}} \big|   \overline{\chi}_{n', \boldsymbol{k}'} \big>  & = \int d \boldsymbol{r} \sqrt{- g(\boldsymbol{r}) }  \,\overline{\chi}^{\dagger}_{n, \boldsymbol{k}}(\boldsymbol{r}) \cdot \overline{\chi}_{n', \boldsymbol{k}'}(\boldsymbol{r}) \nonumber \\
    & = \delta_{n, n'} \,\delta(\boldsymbol{k} - \boldsymbol{k}') \,.
\end{align}
with the representation of the scalar product in a curvilinear reference frame, given by Eq.~\eqref{scalar product in arbitrary metric}. 

It is convenient to introduce the quantity
\begin{align}
    \big(  \Psi \big| \hat{A} \big| \Phi \big) \equiv \int d \boldsymbol{r}\,   \Psi^{\dagger}( \boldsymbol{r} )\, \cdot A(\boldsymbol{r}) \, \Phi(\boldsymbol{r}) \,,
    \label{scalar product in Cartesian metric}
\end{align}
which, as mentioned in the main text, is {\it not} a scalar product in the curvilinear reference frame, which is given instead by Eq.~\eqref{scalar product in arbitrary metric}. 
The quantity in Eq.~\eqref{scalar product in Cartesian metric}, nevertheless, will appear in the following derivation, due to the fact that the accuracy of the theory up to the first order in the strain tensor also requires the expansion of $\sqrt{-g(\boldsymbol{r})}$ in Eq.~\eqref{scalar product in arbitrary metric}.
 
For the purposes of the present derivation, it is convenient to rewrite the Hamiltonian $H = H_{\rm kin} + U_{\rm n} + H_{\rm so}  + U_{\rm ext}$ as $H = H_{0} + H_1   +   U_{\rm ext}$, where
\begin{widetext}
\begin{align}
    H_0     \equiv H_{{\rm kin},0} + U_{{\rm n},0} + H_{{\rm so},0} = - \frac{  \hbar^2 }{2 m}    \frac{\partial^2}{\partial r^{\mu} \partial r^{\mu}} + U_{{\rm n},0}  - \frac{ {\rm i} \hbar^2   }{8 m^2 c^2 }        \Big[       \left( \partial_{\mu} U_{{\rm n},0} \right)  \sigma^{\mu \nu}     \overrightarrow{\partial}_{\nu}     -    \overleftarrow{\partial}_{\nu}     \sigma^{\mu \nu}       \left( \partial_{\mu} U_{{\rm n},0}  \right) \Big]      
    \label{Schroedinger Hamiltonian}
\end{align}
and
\begin{align}
    H_1   \equiv H_{{\rm kin},1} + U_{{\rm n},1} + H_{{\rm so},1} = \underbrace{ - \frac{\hbar^2}{2 m} \overleftarrow{\partial}_{\mu} \left(   \varepsilon^{\nu}_{\mu}  + \varepsilon^{\mu}_{\nu}     
 \right)  \overrightarrow{\partial}_{\nu}    +\varepsilon^{\alpha}_{\beta}  U^{\beta}_{\alpha} }_{ H_{1, {\rm nonrel}}} 
 \underbrace{ - \frac{ {\rm i} \hbar^2   }{8 m^2 c^2 }        \left(   \Sigma^{\nu} \overrightarrow{\partial}_{\nu}     -    \overleftarrow{\partial}_{\nu}    \Sigma^{\nu}  \right) }_{ H_{1, {\rm so}}}  
    \label{strain correction Hamiltonian}
\end{align}    
\end{widetext}
collect the terms which are, respectively, independent of and linear in the strain tensor; $U_{\rm ext}$ is left untouched.

Using the equation
\begin{align}
    \frac{ \partial \overline{\chi}_{n', \boldsymbol{k}'} }{\partial r^{\nu}} \approx   \frac{ \partial \chi_{n', \boldsymbol{k}'} }{\partial r^{\nu}}   - \frac{1}{2}   \frac{ \partial \varepsilon^{\gamma}_{\gamma} }{\partial r^{\nu}}   \chi_{n', \boldsymbol{k}'}     - \frac{1}{2} \varepsilon^{\gamma}_{\gamma}    \frac{ \partial \chi_{n', \boldsymbol{k}'} }{\partial r^{\nu}}    \,, 
\end{align}
and keeping the BvKBCs into account, one obtains:
\begin{align}
    & \big< \overline{\chi}_{n, \boldsymbol{k}} \big|    \hat{H}_{0}    \big| \overline{\chi}_{n', \boldsymbol{k}'} \big> = \int d \boldsymbol{r} \sqrt{- g} \, \overline{\chi}^{\dagger}_{n, \boldsymbol{k}}  H_0 \overline{\chi}_{n', \boldsymbol{k}'}  \nonumber \\
    & \approx  \big(  \chi_{n, \boldsymbol{k}} \big|  \hat{H}_0 \big|   \chi_{n', \boldsymbol{k}'} \big)        - \frac{  \hbar^2 }{4 m}  \int d \boldsymbol{r}  \frac{ \partial \varepsilon^{\gamma}_{\gamma} }{\partial r^{\mu}}    \frac{ \partial \left(   \chi^{\dagger}_{n, \boldsymbol{k}}        \cdot         \chi_{n', \boldsymbol{k}'}        \right)  }{\partial r^{\mu}}         \nonumber \\
    & =  \big(  \chi_{n, \boldsymbol{k}} \big|  \hat{H}_0 \big|   \chi_{n', \boldsymbol{k}'} \big)          + \frac{  \hbar^2 }{4 m}  \int d \boldsymbol{r}  \frac{ \partial^2 \varepsilon^{\gamma}_{\gamma} }{\partial r^{\mu} \partial r^{\mu} }   \,    \left(   \chi^{\dagger}_{n, \boldsymbol{k}}        \cdot         \chi_{n', \boldsymbol{k}'}        \right)   \nonumber \\
    & = \big(  \chi_{n, \boldsymbol{k}} \big|  \hat{H}_0 \big|   \chi_{n', \boldsymbol{k}'} \big)          + \frac{  \hbar^2 }{4 m}  \big(  \chi_{n, \boldsymbol{k}} \big| \left( \nabla^2 {\rm tr} \varepsilon \right) \big|   \chi_{n', \boldsymbol{k}'} \big) \,,
\end{align}
accurately to the first order in the strain tensor. Then, the equation
\begin{align}
    \big< \overline{\chi}_{n, \boldsymbol{k}} \big|      \hat{U}_{\rm ext}    \big| \overline{\chi}_{n', \boldsymbol{k}'} \big> = \big( \chi_{n, \boldsymbol{k}} \big|      \hat{U}_{\rm ext}    \big| \chi_{n', \boldsymbol{k}'} \big) 
\end{align}
holds exactly, because $\hat{U}_{\rm ext} $ is a function of position only, and
\begin{align}
    \big< \overline{\chi}_{n, \boldsymbol{k}} \big|      \hat{H}_1    \big| \overline{\chi}_{n', \boldsymbol{k}'} \big> \approx \big( \chi_{n, \boldsymbol{k}} \big|      \hat{H}_1    \big| \chi_{n', \boldsymbol{k}'} \big) \,,
\end{align}
because $\hat{H}_1$ is of order 1 in the components of the strain tensor.

As a result, one has that:
\begin{align}
    \big< \overline{\chi}_{n, \boldsymbol{k}} \big|    \hat{H}    \big| \overline{\chi}_{n', \boldsymbol{k}'} \big> & \approx \big(  \chi_{n, \boldsymbol{k}} \big|  \hat{H}_0 \big|   \chi_{n', \boldsymbol{k}'} \big)         + \big( \chi_{n, \boldsymbol{k}} \big|      \hat{U}_{\rm ext}    \big| \chi_{n', \boldsymbol{k}'} \big)  \nonumber \\
    & \quad + \frac{  \hbar^2 }{4 m}  \big(  \chi_{n, \boldsymbol{k}} \big| \left( \nabla^2 {\rm tr} \varepsilon \right) \big|   \chi_{n', \boldsymbol{k}'} \big)  \nonumber \\
    & \quad + \big( \chi_{n, \boldsymbol{k}} \big|      \hat{H}_1    \big| \chi_{n', \boldsymbol{k}'} \big)     \,.
    \label{basis for LK}
\end{align}
The detailed evaluation of the matrix elements appearing in the right-hand side of Eq.~\eqref{basis for LK} is presented in Sections I and II of the SM \cite{SuppMat}.

\section{Envelope-function Hamiltonian}
\label{app: envelope}

Since the transformed Hamiltonian, obtained after the manifold decoupling, is block-diagonal, its eigenstates are combinations of the basis states belonging to a single block. For the low-energy block, an eigenstate is written as
\begin{align}
    \big| \phi \big> \equiv \sum_{n \leq N} \int_{1 \rm BZ} d \boldsymbol{k} \, \mathcal{C}_n(\boldsymbol{k}) \big| \overline{\chi}_{n, \boldsymbol{k}} \big> \,,
    \label{block phi}
\end{align}
where the coefficients $\mathcal{C}_n(\boldsymbol{k})$ satisfy
\begin{align}
   \sum_{n' \leq N} \int_{\rm 1 BZ} d \boldsymbol{k}' \, \mathcal{H}^{(N)}_{n, n'}(\boldsymbol{k}, \boldsymbol{k}') \, \mathcal{C}_{n'}(\boldsymbol{k}') = E \mathcal{C}_n(\boldsymbol{k}) \,.
   \label{secular equation N manifold}
\end{align}

The slowly-varying \emph{envelope functions} are defined as
\begin{align}
    F_n(\boldsymbol{r}) \equiv \int_{1 \rm BZ} d \boldsymbol{k} \,{\rm e}^{{\rm i} \boldsymbol{k} \cdot \boldsymbol{r}} \mathcal{C}_{n}(\boldsymbol{k}) \,, 
    \label{envelope functions}
\end{align} 
and they allow to write the spinorial wave function $\big< \boldsymbol{r} \big| \phi \big>$ in the form given by Eq.~\eqref{wfc curvilinear coordinates} of the main text. Equation \eqref{secular equation N manifold} is rewritten in terms of the envelope functions as
\begin{align}
   & \sum_{n' \leq N} \int_{\rm 1 BZ} d \boldsymbol{k} \, {\rm e}^{{\rm i} \boldsymbol{k} \cdot \boldsymbol{r}}   \int_{\rm 1 BZ} d \boldsymbol{k}' \, \mathcal{H}^{(N)}_{n, n'}(\boldsymbol{k}, \boldsymbol{k}') \, \mathcal{C}_{n'}(\boldsymbol{k}') \nonumber \\
   &  = E F_n(\boldsymbol{r}) \,.
   \label{secular equation envelope functions}
\end{align}

To simplify the left-hand side of Eq.~\eqref{secular equation envelope functions}, we distinguish three types of matrix elements of the Hamiltonian:
\begin{itemize}
    \item those having the form $\delta(\boldsymbol{k} - \boldsymbol{k}') \Lambda_{n,n'}(\boldsymbol{k})$: these give
    \begin{align}
        \sum_{n' \leq N} \Lambda_{n,n'}(- {\rm i} \nabla) F_{n'}(\boldsymbol{r}) \,,
    \end{align}
    analogously to standard envelope-function theories;
    \item those having the form $\widetilde{\mathcal{U}}_{n, n'}(\boldsymbol{k} - \boldsymbol{k}')$: these include the external potential and formally analogous terms, for which the standard treatment is applicable; under the assumption that the envelope functions are slowly varying, they contribute terms
    \begin{align}
        \sum_{n' \leq N} \mathcal{U}_{n, n'}(\boldsymbol{r} ) F_{n'}(\boldsymbol{r}) \,;
    \end{align}
    \item those having the form $\widetilde{\varepsilon}^{\, \mu}_{\nu}(\boldsymbol{k} - \boldsymbol{k}')    \left[ X^{\nu}_{\mu}(\boldsymbol{k}, \boldsymbol{k}') \right]_{n, n'}$. This is a new category of terms, which do not map onto those related to homogeneous strain, because the Fourier transform of the strain tensor is not a Dirac delta. 
\end{itemize}

The contributions to the left-hand side of Eq.~\eqref{secular equation envelope functions} which include the formally new terms are written as
\begin{align}
   & \sum_{n' \leq N} \int_{\rm 1 BZ} d \boldsymbol{k} \, {\rm e}^{{\rm i} \boldsymbol{k} \cdot \boldsymbol{r}}  \int_{\rm 1 BZ} d \boldsymbol{k}' \, \widetilde{\varepsilon}^{\, \mu}_{\nu}(\boldsymbol{k} - \boldsymbol{k}')    \left[ X^{\nu}_{\mu}(\boldsymbol{k}, \boldsymbol{k}') \right]_{n, n'} \nonumber \\
   & \quad \times \mathcal{C}_{n'}(\boldsymbol{k}')  \nonumber \\
   & = \sum_{n' \leq N} \int d \boldsymbol{r}'  \varepsilon^{\, \mu}_{\nu}(\boldsymbol{r}')  \frac{1}{(2 \pi)^3} 
 \int_{\rm 1 BZ} d \boldsymbol{k} \, {\rm e}^{{\rm i} \boldsymbol{k} \cdot \left( \boldsymbol{r} - \boldsymbol{r}' \right) }    \nonumber \\
 & \quad \times \int_{\rm 1 BZ} d \boldsymbol{k}'   {\rm e}^{  {\rm i}  \boldsymbol{k}'   \cdot \boldsymbol{r}'} \Big\{ \left( \mathcal{D}_{\mu}^{\nu} \right)_{n, n'}       + k^{\alpha} \left( \mathcal{L}_{\alpha; \mu}^{\nu} \right)_{n, n'}  \nonumber \\
 & \quad + k'^{\alpha} \left( \mathcal{L}_{\alpha; \mu}^{*\nu} \right)_{n', n} + k^{\alpha} k'^{\beta}  \left( \mathcal{Q}_{\alpha \beta; \mu}^{\nu} \right)_{n, n'}  \Big\}   \, \mathcal{C}_{n'}(\boldsymbol{k}') \,.
 \label{new terms}
\end{align}
The first contribution to the right-hand side of Eq.~\eqref{new terms} is
\begin{align}
   &   \sum_{n' \leq N} \left( \mathcal{D}_{\mu}^{\nu} \right)_{n, n'}  \int d \boldsymbol{r}'  \varepsilon^{\, \mu}_{\nu}(\boldsymbol{r}')  \frac{1}{(2 \pi)^3} 
 \int_{\rm 1 BZ} d \boldsymbol{k} \, {\rm e}^{{\rm i} \boldsymbol{k} \cdot \left( \boldsymbol{r} - \boldsymbol{r}' \right) } \nonumber \\
 & \times \int_{\rm 1 BZ} d \boldsymbol{k}'   {\rm e}^{  {\rm i}  \boldsymbol{k}'   \cdot \boldsymbol{r}'}             \mathcal{C}_{n'}(\boldsymbol{k}')  \nonumber \\
 & \approx   \varepsilon^{\, \mu}_{\nu}(\boldsymbol{r})  \sum_{n' \leq N} \left( \mathcal{D}_{\mu}^{\nu} \right)_{n, n'}        F_{n'} (\boldsymbol{r}) \,,
\end{align}
the second contribution is
\begin{align}
&   \sum_{n' \leq N} \left( \mathcal{L}_{\alpha; \mu}^{\nu} \right)_{n, n'}  \int d \boldsymbol{r}'  \varepsilon^{\, \mu}_{\nu}(\boldsymbol{r}')  \frac{1}{(2 \pi)^3} 
 \int_{\rm 1 BZ} d \boldsymbol{k} \, {\rm e}^{{\rm i} \boldsymbol{k} \cdot \left( \boldsymbol{r} - \boldsymbol{r}' \right) }  k^{\alpha} \nonumber \\
 & \times \int_{\rm 1 BZ} d \boldsymbol{k}'   {\rm e}^{  {\rm i}  \boldsymbol{k}'   \cdot \boldsymbol{r}'}     \mathcal{C}_{n'}(\boldsymbol{k}') \nonumber \\
& \approx - {\rm i}  \sum_{n' \leq N} \left( \mathcal{L}_{\alpha; \mu}^{\nu} \right)_{n, n'}  \int d \boldsymbol{r}'  \varepsilon^{\, \mu}_{\nu}(\boldsymbol{r}') F_{n'}(\boldsymbol{r}')   \frac{\partial \delta \left( \boldsymbol{r} - \boldsymbol{r}' \right)}{\partial r^{\alpha}}  \nonumber \\
& = - {\rm i}  \sum_{n' \leq N} \left( \mathcal{L}_{\alpha; \mu}^{\nu} \right)_{n, n'}       \frac{\partial  \left[ \varepsilon^{\, \mu}_{\nu}(\boldsymbol{r}) F_{n'}(\boldsymbol{r}) \right]  }{\partial r^{\alpha}}     \,,
\end{align}
the third contribution is
\begin{align}
&   \sum_{n' \leq N} \left( \mathcal{L}_{\alpha; \mu}^{*\nu} \right)_{n', n}  \int d \boldsymbol{r}'  \varepsilon^{\, \mu}_{\nu}(\boldsymbol{r}')  \frac{1}{(2 \pi)^3} 
 \int_{\rm 1 BZ} d \boldsymbol{k} \, {\rm e}^{{\rm i} \boldsymbol{k} \cdot \left( \boldsymbol{r} - \boldsymbol{r}' \right) } \nonumber \\
 & \times \int_{\rm 1 BZ} d \boldsymbol{k}'   {\rm e}^{  {\rm i}  \boldsymbol{k}'   \cdot \boldsymbol{r}'}       k'^{\alpha}      \, \mathcal{C}_{n'}(\boldsymbol{k}') \nonumber \\
 & \approx - {\rm i} \sum_{n' \leq N} \left( \mathcal{L}_{\alpha; \mu}^{*\nu} \right)_{n', n}    \varepsilon^{\, \mu}_{\nu}(\boldsymbol{r})  \frac{\partial F_{n'}(\boldsymbol{r}) }{\partial r^{\alpha}}      \,,
\end{align}
and the fourth contribution is
\begin{align}
&   \sum_{n' \leq N} \left( \mathcal{Q}_{\alpha \beta; \mu}^{\nu} \right)_{n, n'}  \int d \boldsymbol{r}'  \varepsilon^{\, \mu}_{\nu}(\boldsymbol{r}')  \frac{1}{(2 \pi)^3} 
 \int_{\rm 1 BZ} d \boldsymbol{k} \, {\rm e}^{{\rm i} \boldsymbol{k} \cdot \left( \boldsymbol{r} - \boldsymbol{r}' \right) }  k^{\alpha} \nonumber \\
 & \times \int_{\rm 1 BZ} d \boldsymbol{k}'   {\rm e}^{  {\rm i}  \boldsymbol{k}'   \cdot \boldsymbol{r}'}       k'^{\beta}     \mathcal{C}_{n'}(\boldsymbol{k}') \nonumber \\
& \approx -   \sum_{n' \leq N} \left( \mathcal{Q}_{\alpha \beta; \mu}^{\nu} \right)_{n, n'}  \int d \boldsymbol{r}'  \varepsilon^{\, \mu}_{\nu}(\boldsymbol{r}')   \frac{\partial  F_{n'}(\boldsymbol{r}') }{\partial r'^{\beta}} \frac{\partial \delta\left( \boldsymbol{r} - \boldsymbol{r}' \right)  }{\partial r^{\alpha}}  \nonumber \\
& = -   \sum_{n' \leq N} \left( \mathcal{Q}_{\alpha \beta; \mu}^{\nu} \right)_{n, n'} \frac{\partial   }{\partial r^{\alpha}}  \left[ \varepsilon^{\, \mu}_{\nu}(\boldsymbol{r})   \frac{\partial  F_{n'}(\boldsymbol{r}) }{\partial r^{\beta}}  \right]     \,.
\end{align} 
In deriving the expressions above, we have used
\begin{align}
  \frac{1}{(2 \pi)^3}  \int_{\rm 1 BZ} d \boldsymbol{k} \, {\rm e}^{{\rm i} \boldsymbol{k} \cdot \left( \boldsymbol{r} - \boldsymbol{r}' \right) } \approx \delta\left( \boldsymbol{r} - \boldsymbol{r}' \right) \,,
\end{align}
which is an approximate relation, only valid when this quantity is multiplied by a slowly-varying spatial function, such as envelope functions and components of the strain tensor.

Collecting all terms, one obtains 
\begin{align}
    \sum_{n' \leq N} \left[ \left( \hat{\mathcal{H}}^{(0)}_{\rm EF} \right)_{n,n'} + \left( \hat{\mathcal{H}}^{(1)}_{\rm EF} \right)_{n,n'} \right] F_{n'}(\boldsymbol{r})     
  = E F_n(\boldsymbol{r}) \,,
\end{align}
where
\begin{align}
    \left( \hat{\mathcal{H}}^{(0)}_{\rm EF} \right)_{n,n'} & \equiv \left[  E_n(\boldsymbol{0}) + \frac{ \hbar^2 \hat{\boldsymbol{k}}^2 }{2 m} + U_{\rm ext}(\boldsymbol{r})  \right] \delta_{n, n'} \nonumber \\
    & \quad + \frac{\hbar \pi^{\alpha}_{n, n'}}{m}       \hat{ k }_{\alpha} +  \frac{\hbar^2 \Pi^{\alpha \beta}_{n, n'} }{m^2} \hat{k}_{\alpha} \hat{k}_{\beta}   
\end{align}
is formally the same as the standard $\boldsymbol{k} \cdot \boldsymbol{p}$ Hamiltonian \cite{VoonWillatzen}, but with the $\hat{k}_{\alpha}$ operators defined in curvilinear coordinates, and $\left( \hat{\mathcal{H}}^{(1)}_{\rm EF} \right)_{n,n'}$ is the strain-dependent term defined in Eq.~\eqref{secular equation envelope functions final}.

\section{Numerical estimates}
\label{app: numerics}

\subsection{Strain model in a MOSFET}

Here we provide quantitative estimates for the new Hamiltonian terms that depend on the inhomogeneous strain tensor and its derivatives. For simplicity, we refer to the approximate Hamiltonian of Eq.~\eqref{6x6 matrix} and we apply it to describe a model silicon nanostructure. We consider a system similar to the one described in Ref.~\cite{Bonera13}, consisting of a MOSFET with two SiGe stressors placed, respectively, above the source and the drain gates. An analytical model that approximately describes the stress tensor in such a system can be derived by solving the elasticity problem of a localized force on a semi-infinite plate \cite{Bonera13, TimoshenkoGoodier}. 

From the stress tensor, one should then obtain the strain tensor $\varepsilon^{\alpha}_{\beta}(\boldsymbol{r})$ by applying the compliance relations \cite{Sverdlov}, and the displacement $u^{\alpha}(\boldsymbol{r})$ by integrating the strain tensor according to Eq.~\eqref{relation displacement and strain}, taking e.g. $\boldsymbol{r}_0 = \boldsymbol{0}$ and $\boldsymbol{u}(\boldsymbol{0}) = \boldsymbol{0}$ (which is equivalent to fixing the origin of the laboratory reference frame). However, the expression of the stress tensor given in Ref.~\cite{Bonera13} is notoriously an approximate one, which does not satisfy the compatibility equations exactly \cite{TimoshenkoGoodier}. A consequence of this fact is that there exists no function $u^{\alpha}(\boldsymbol{r})$ such that the strain tensor derived from the given stress (let us call it $\varepsilon_{\rm approx.}$) satisfies Eq.~\eqref{definition of strain}. 

An equivalent alternative formulation of this statement is that the line integral in Eq.~\eqref{relation displacement and strain}, when evaluated on $\varepsilon_{\rm approx.}$, depends on the path connecting $\boldsymbol{r}_0$ to $\boldsymbol{r}$. This can be verified by comparing the results obtained with the following paths [with $\boldsymbol{r} \equiv (x,y,z)$]:
\begin{align}
    & {\rm Path \, 1:} \quad (0,0,0) \rightarrow (x,0,0)   \rightarrow (x,y,0)  \rightarrow (x,y,z) \,, \nonumber \\
    & {\rm Path \, 2:} \quad (0,0,0) \rightarrow (0,0,z)   \rightarrow (0,y,z)  \rightarrow (x,y,z) \,.
\end{align}
We notice, however, that the displacements obtained by using these two paths are nearly identical almost everywhere. The discrepancies occur on small domains close to the stressors, where the simplifying assumptions beyond the analytical solution for the stress tensor apparently imply some non-physical behavior if Path 1 is used. 

Therefore, we assume that the displacement obtained with Path 2 is the physically correct one, and we adopt its expression for the function $\boldsymbol{u}(\boldsymbol{r})$. Inverting our perspective, we now take this as our starting point, and we obtain the strain tensor $\varepsilon^{\alpha}_{\beta}(\boldsymbol{r})$ analytically by applying Eq.~\eqref{definition of strain}. The strain tensor obtained in this way, by construction, is uniquely defined and obviously satisfies the compatibility conditions.

The displacement field that we adopt is then given by:
\begin{widetext}
    \begin{align}
    u^x(\boldsymbol{r}) & = \frac{  \sigma_0 t}{2 \pi} \sum_{i = 1}^4 s_i \Bigg\{     \left(  - s_{11} + s_{12}  + \frac{s_{44}}{2} \right) \frac{    z^2  }{ x_i^2+z^2  } 
    + \left(  s_{11} - s_{12}  + \frac{s_{44}}{2} \right) \frac{     z^2  }{  (x-x_i)^2+z^2 }   \nonumber \\
    & \quad 
      + \left(  s_{11} + s_{12}  + \frac{s_{44}}{2} \right) \ln \left( \frac{(x-x_i)^2+z^2}{x_i^2 + z^2 } \right)    \Bigg\} \,, \nonumber \\
    u^y(\boldsymbol{r}) & = 0 \,, \nonumber \\
    u^z(\boldsymbol{r}) & = \frac{   \sigma_0 t}{\pi} \sum_{i = 1}^4 s_i \Bigg\{  
    \left( s_{11} -  s_{12} - \frac{s_{44}}{2} \right) \frac{x_i z}{x_i^2+z^2} 
    - \frac{s_{44}}{2} \frac{(x-x_i) z}{(x-x_i)^2+z^2}
    - \left( s_{11} + s_{12} \right) \arctan\left(\frac{z}{x_i}\right)      \nonumber \\
    & \quad 
    + \frac{s_{44}}{2} \arctan\left(\frac{x_i}{z}\right) 
    - \frac{s_{44}}{2} \arctan\left(  \frac{x_i-x}{z}\right)   \Bigg\} \,.
    \label{adopted u}
\end{align} 
\end{widetext}
Here: $s_{11}$, $s_{12}$ and $s_{44}$ are the compliance constants of Si \cite{Sverdlov}; $\lbrace x_i; \, i = 1,2,3,4 \rbrace$ is the set of sidewall positions defining the extension of the stressors; generalizing Ref.~\cite{Bonera13}, we take
\begin{align}
    x_1 = -w - \frac{d}{2} \,, \quad x_2 =   - \frac{d}{2} \,, \quad x_3 =   \frac{d}{2} \,, \quad x_4 =  w + \frac{d}{2} \,,
\end{align}
where $w$ is the stressors' width (along the $x$ direction) and $d$ is their separation ($d$ is larger than the MOSFET channel length); $s_i$ are signs that account for the orientations of the forces applied by the stressors (taking $s_1 = s_3 = +1$ and $s_2 = s_4 = -1$, one reproduces the configuration of Ref.~\cite{Bonera13}); $t$ is the thickness of the stressors in the $z$ direction, and $\sigma_0$ is the biaxial stress parameter in the SiGe stressors.

\subsection{Main new terms due to the inhomogeneity of strain}

We focus on the approximate envelope-function Hamiltonian given in Eq.~\eqref{6x6 matrix} and discuss the three main new terms resulting from our formulation, all of which are diagonal in the band index:
\begin{align}
 & \hat{T}_1 \equiv   \frac{\hbar^2}{4 m} \left[ \nabla^2 \varepsilon^{\mu}_{\mu}(\boldsymbol{r}) \right] \,, \quad \hat{T}_2 \equiv - \frac{\hbar^2  }{m}   
     \varepsilon^{\rm sym }_{\alpha \beta}(\boldsymbol{r})       \hat{k}_{\alpha} \hat{k}_{\beta}      \,, \nonumber \\
     & \hat{T}_3 \equiv  \frac{\hbar^2  }{m} \left[ \partial_{\alpha}  \varepsilon^{\rm sym}_{\alpha \beta}(\boldsymbol{r}) \right]  {\rm i} \, \hat{k}_{\beta} 
     \label{new terms T1 T2 T3}
       \,. 
\end{align}

We start from the first term, $\hat{T}_1$. The trace of the strain tensor is proportional to the trace of the stress tensor, and it can be shown \cite{TimoshenkoGoodier} that the laplacian of the latter vanishes in an \emph{isotropic} system. Therefore, in such systems $\hat{T}_1$ vanishes identically. However, crystals are not isotropic, and assuming that they are is an approximation that might fail on small spatial scales such as those pertaining to nanostructures (the anisotropy of crystals is reflected in the values of the compliance parameters, which connect stress and strain tensors). Overall, $\hat{T}_1$ is a function of coordinates only, which appears on the diagonal of the $\boldsymbol{k} \cdot \boldsymbol{p}$ matrix. Therefore, it represents a correction to the slow-varying confinement potential.   

The term $\hat{T}_2$ is a quadratic function of the momentum operator acting on the envelope functions; it can be viewed as a spatially-dependent correction to the effective-mass terms, where the spatial dependence is due to the inhomogeneity of the strain tensor. This is the only term among those in Eq.~\eqref{new terms T1 T2 T3} that survives if the strain is homogeneous; however, it has been ignored so far. 

The term $\hat{T}_3$ has a completely new form, being linear in the momentum operator. It is not analogous to a spin-orbit term, since it does not couple different bands.

\subsection{Numerical values} 

\begin{figure}
\centering
\includegraphics[width=0.35\textwidth]{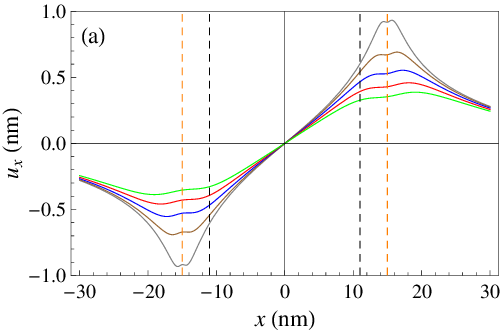}
\quad 
\includegraphics[width=0.35\textwidth]{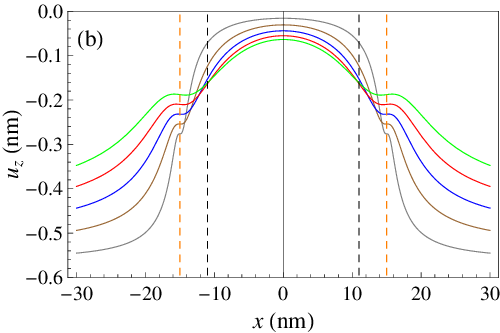}
\caption{Components (a) $u^x$ and (b) $u^z$ of the displacement from Eq.~\eqref{adopted u}, as functions of $x$ and for selected values of $z =$ $1$ nm (gray), $2$ nm (brown), $3$ nm (blue), $4$ nm (red), $5$ nm (green). Parameters ($\sigma_0 = -1.9$ GPa, $ t = 50$ nm, $w = 180$ nm) are taken from Ref.~\cite{Bonera13}, except for $d=30$ nm. Vertical dashed lines correspond to the extremities of the channel at $x = \pm 11$ nm (black) and to the internal borders of the stressors at $x = \pm 15$ nm (orange).}
\label{fig disp}
\end{figure} 

We here present numerical estimates related to the quantities $\hat{T}_1$, $\hat{T}_2$ and $\hat{T}_3$. For illustrative purposes, we use the experimental values given in Ref.~\cite{Bonera13} to define the stressors ($\sigma_0 = -1.9$ GPa, $w = 180$ nm, $t = 50$ nm), but we assume that the distance $d$ between them can be changed. In particular, we apply this analysis to a hypothetical device based on the 22nm-FDSOI technology \cite{Carter16}, and we take $d=30$ nm. The resulting displacement field is displayed in Fig.~\ref{fig disp}. The following discussion is not meant to be exhaustive of all possible cases; it rather illustrates the main physical features of the new terms. Their actual values and the extent of their impact on the physical properties of the confined hole states are strongly device-dependent, and a comprehensive analysis of the parameter space is beyond the scope of the present work.

\begin{figure}
\centering
\includegraphics[width=0.35\textwidth]{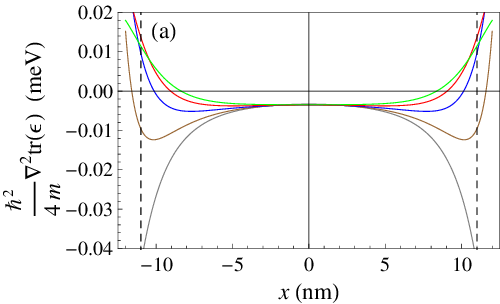}
\quad 
\includegraphics[width=0.35\textwidth]{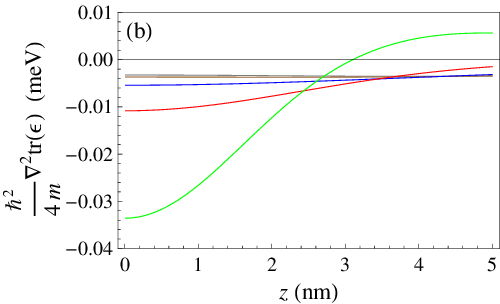}
\caption{Quantity $T_1(\boldsymbol{r})$ (a) as a function of $x$ for selected values of $z$ (same colors and conventions as in Fig.~\ref{fig disp}), and (b) as a function of $z$ for selected values of $x = 0$ nm (gray), $2.5$ nm (brown), $5$ nm (blue), $7.5$ nm (red), and $10$ nm (green).}
\label{fig T1}
\end{figure}

In Fig.~\ref{fig T1} we plot $T_1(\boldsymbol{r})$ as a function of position. As discussed in the main text, this quantity should be thought of as an effective correction to the confinement potential. As can be seen from Fig.~\ref{fig T1}(a), for values of $z$ close to the top of the channel, this correction develops attractive (repulsive) pockets for electrons (holes), close to the ends of the channel along the $x$ direction. In certain geometries, this might be an unwanted source of perturbation for the confinement, especially in the electron case, where the lateral attractive pockets compete with the confinement at the center of the channel. In the particular device discussed here, however, such effect seems to be minor, due to the small values achieved by $T_1$ over the channel and its near-uniformity close to the center of the channel.

\begin{figure}
\centering
\includegraphics[width=0.35\textwidth]{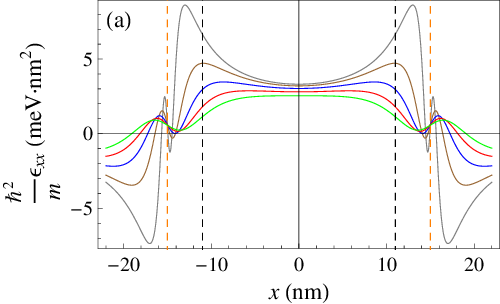}
\quad 
\includegraphics[width=0.35\textwidth]{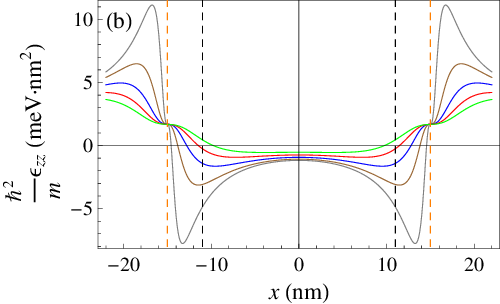}
\quad 
\includegraphics[width=0.35\textwidth]{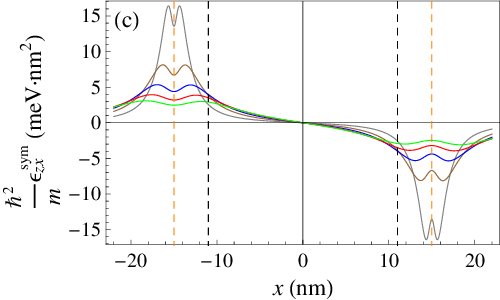}
\caption{Prefactors of the quadratic forms of the momentum components appearing in $\hat{T}_2$ as functions of $x$, for selected values of $z$ (same colors and conventions as in Fig.~\ref{fig disp}). (a) coefficient of $\hat{k}_x^2$, (b) coefficient of $\hat{k}_z^2$, (c) coefficient of $\hat{k}_x \hat{k}_z$.}
\label{fig T2}
\end{figure}

The operator $\hat{T}_2$ is determined by the coefficients of the quadratic forms $\hat{k}_{\alpha} \hat{k}_{\beta}$, which, apart from a multiplicative constant, are the symmetric components of the strain tensor. They are shown in Fig.~\ref{fig T2}. For the considered device, the diagonal terms $(xx)$ and $(zz)$ are even functions of $x$, and they reach values of the order of several meV $\cdot$ nm$^2$ close to the center of the channel, while the $(zx)$ term is an odd function of $x$, and its values in the relevant region are smaller. For comparison, the coefficients of $\hat{k}_x^2$ and $\hat{k}_z^2$ in the $\boldsymbol{k} \cdot \boldsymbol{p}$ Hamiltonian for heavy and light holes in silicon are:
\begin{align}
    & \frac{\hbar^2}{2 m_{x, {\rm H} }}   = 176.174 \,\, \text{meV}\cdot \text{nm}^2 \,, \nonumber \\
    & \frac{\hbar^2}{2 m_{x, {\rm L} }}   = 150.342 \,\, \text{meV}\cdot \text{nm}^2 \,, \nonumber \\
    &  \frac{\hbar^2}{2 m_{z, {\rm H} }}   = 137.426 \,\, \text{meV}\cdot \text{nm}^2 \,, \nonumber \\
    & \frac{\hbar^2}{2 m_{z, {\rm L} }}   = 189.089 \,\, \text{meV}\cdot \text{nm}^2 \,,
\end{align}
where $m_{\alpha, h}$ is the effective mass along direction $\alpha$, and $h = H / L$ denotes heavy/light holes. Therefore, the $\hat{T}_2$ term produces a sizeable, spatially-dependent correction to the effective masses, which for the considered device can be of the order of $\approx 1 \%$ to $2 \%$, depending on the hole species and on the direction.

\begin{figure}
\centering
\includegraphics[width=0.35\textwidth]{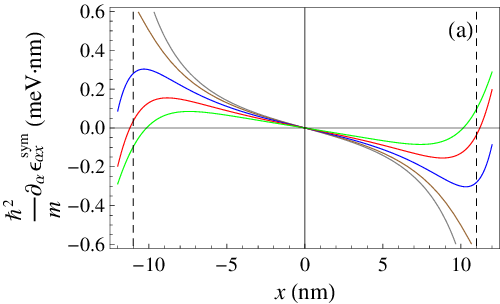}
\quad 
\includegraphics[width=0.35\textwidth]{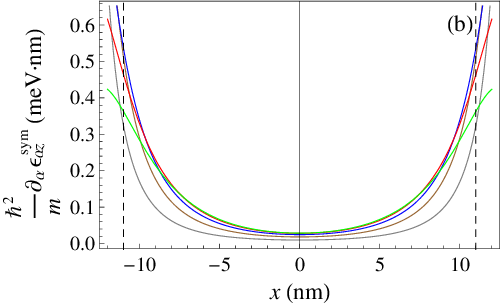}
\caption{Prefactors of the linear forms of the momentum components appearing in $\hat{T}_3$ as functions of $x$, for selected values of $z$ (same colors and conventions as in Fig.~\ref{fig disp}). (a) coefficient of $\hat{k}_x$, (b) coefficient of $\hat{k}_z$.}  
\label{fig T3}
\end{figure}

Finally, the term $\hat{T}_3$ is determined by the spatially-dependent coefficients multiplying the components of the momentum operator. For the case at hand, these are displayed in Fig.~\ref{fig T3}. The considered geometry produces coefficients of $\hat{k}_x$ and $\hat{k}_z$ that are, respectively, an odd and an even function of $x$. Therefore, in this particular case, the $\hat{k}_x$ term likely has a minor impact on the states of the confined holes; the $\hat{k}_z$ term, instead, is associated with characteristic energies of the order of $\left< \frac{\hbar^2}{m} \partial_{\alpha} \varepsilon_{\alpha z}^{\rm sym} \right> \hat{k}_z \approx  \overline{ \frac{\hbar^2}{m} \partial_{\alpha} \varepsilon_{\alpha z}^{\rm sym} }    \frac{1}{L_z}$, where $L_z$ is a typical confinement length along the $z$ direction, and $\overline{ \frac{\hbar^2}{m} \partial_{\alpha} \varepsilon_{\alpha z}^{\rm sym} }$ is an average value of the coefficient depicted in Fig.~\ref{fig T3}(b) close to the center of the channel. Taking for the latter a representative value of $0.1$ meV$\cdot$nm, and $L_z \approx 10$ nm or $5$ nm (as in currently available or downscaled MOSFETs, respectively \cite{Bellentani21a}), one obtains energy scales of the order of $\approx 0.01 \div 0.02$ meV, corresponding to effective frequencies of $\approx 2.4 \div 4.8$ GHz.

\begin{widetext}

\section*{SUPPLEMENTARY MATERIAL}


In this Supplementary Material, we provide technical details related to the derivation of the results presented in the main text. In particular, in Section I we derive the matrix elements of the Hamiltonian in curvilinear coordinates on the basis of modified Luttinger-Kohn states, and in Section II we detail the procedure for decoupling the manifold of interest from the remote bands.

In the following, equations specified by a number preceded by S [e.g., Eq.~(S1)] are those of the present Supplementary Material, while other specifications [e.g., Eq.~(10) or Eq.~(E9)] refer to the main text or its Appendices.

\section*{I. Evaluation of the matrix elements needed for the secular equation}
  
We now evaluate the four terms in the left-hand side of Eq.~(E9)
one by one. In the following, these definitions will be used:
\begin{align}
    u^{\sigma}_{n, \boldsymbol{0}}(\boldsymbol{r}) \equiv \sum_{\boldsymbol{G}} {\rm e}^{{\rm i} \boldsymbol{G} \cdot \boldsymbol{r}} \widetilde{u}^{\sigma}_{n, \boldsymbol{0}}(\boldsymbol{G}) \,, \tag{S1}
\end{align}
with the normalization
\begin{align}
    (2 \pi)^3 \sum_{\sigma} \sum_{\boldsymbol{G}}   \widetilde{u}_{n, \boldsymbol{0}}^{\sigma *}(\boldsymbol{G}) \, \widetilde{u}_{n', \boldsymbol{0}}^{\sigma}(\boldsymbol{G}) = \delta_{n, n'} \,, \tag{S2}
\end{align}
and
\begin{align}
    U_{{\rm n}, 0}(\boldsymbol{r}) \equiv \sum_{\boldsymbol{G}} {\rm e}^{{\rm i} \boldsymbol{G} \cdot \boldsymbol{r}} \widetilde{U}_{{\rm n}, 0}(\boldsymbol{G})  \,. \tag{S3}
\end{align}

It is also convenient to introduce the Fourier transform of the strain tensor (on the infinite $\boldsymbol{q}$ domain),
\begin{align}
    \varepsilon^{\alpha}_{\beta}(\boldsymbol{r}) \equiv   \int d \boldsymbol{q} \, \widetilde{\varepsilon}^{\, \alpha}_{\beta}(\boldsymbol{q}) \, {\rm e}^{{\rm i} \boldsymbol{q} \cdot \boldsymbol{r}}  \quad \Leftrightarrow \quad 
    \widetilde{\varepsilon}^{\, \alpha}_{\beta}(\boldsymbol{q}) \equiv  \frac{1}{(2\pi)^3} \int d \boldsymbol{r} \, \varepsilon^{\alpha}_{\beta}(\boldsymbol{r}) \, {\rm e}^{-{\rm i} \boldsymbol{q} \cdot \boldsymbol{r}}  \,, \tag{S4} 
    \label{strain Fourier}
\end{align}
where $\widetilde{\varepsilon}^{\,\alpha}_{\beta}(- \boldsymbol{q}) = \left[ \widetilde{\varepsilon}^{\,\alpha}_{\beta}(\boldsymbol{q}) \right]^*$, because the strain tensor is real. Homogeneous strain is obtained as a particular case, by setting $\varepsilon^{\nu}_{\lambda}(\boldsymbol{r}) = \varepsilon^{\nu}_{\lambda}   \Leftrightarrow   \widetilde{\varepsilon}^{\, \nu}_{\lambda}(\boldsymbol{q}) = \varepsilon^{\nu}_{\lambda} \, \delta(\boldsymbol{q}) $ .

\subsection{Term due to $\hat{H}_0$}

The first term in the right-hand side of Eq.~(E9)
is evaluated using standard techniques \cite{VoonWillatzen, Luttinger55a}, since it is formally the same as the one arising in $\boldsymbol{k} \cdot \boldsymbol{p}$ theories for unstrained systems in Cartesian coordinates:
\begin{align}
    \big( \chi_{n, \boldsymbol{k}} \big|      \hat{H}_{0}      \big| \chi_{n', \boldsymbol{k}'} \big) & = \int d \boldsymbol{r}   \,  \chi^{\dagger}_{n, \boldsymbol{k}}(\boldsymbol{r})  \left\{ \frac{- \hbar^2 \nabla^2_{\boldsymbol{r}}}{2 m}  +    U_{{\rm n}, 0}(\boldsymbol{r} )  - {\rm i} \, \frac{\hbar^2}{4 m^2 c^2 }    \boldsymbol{\sigma}  \cdot \Big[\left[ \nabla_{\boldsymbol{r}} U_{{\rm n}, 0}(\boldsymbol{r})  \right] \times \nabla_{\boldsymbol{r}} \Big]   \right\}  \chi_{n', \boldsymbol{k}'}(\boldsymbol{r})   \nonumber \\
    & =   \delta(\boldsymbol{k} - \boldsymbol{k}') \left\{ \delta_{n, n'}  \, \left[ E_n(\boldsymbol{0}) + \frac{ \hbar^2 \boldsymbol{k}^2 }{2 m} \right]   + \frac{\hbar }{m} \boldsymbol{k} \cdot \boldsymbol{\pi}_{n, n'}  \right\}   \,. \tag{S5}
    \label{H0 matrix element final}
\end{align}
Here, $E_n(\boldsymbol{0})$ is the energy eigenvalue of band $n$ at the expansion point (here taken to be $\boldsymbol{\Gamma} \equiv \boldsymbol{0}$), and $\boldsymbol{\pi}_{n, n'}$ is defined according to the following equation:
\begin{align}
   \delta\left( \boldsymbol{k} - \boldsymbol{k}' \right) \boldsymbol{\pi}_{n, n'}  \equiv  \int d \boldsymbol{r}  \,  {\rm e}^{- {\rm i} \left( \boldsymbol{k} - \boldsymbol{k}' \right) \cdot \boldsymbol{r}} u^{\dagger}_{n, \boldsymbol{0}}(\boldsymbol{r})    \boldsymbol{\pi}_{\boldsymbol{r}} u_{n', \boldsymbol{0}}(\boldsymbol{r})  \,, \tag{S6}
\label{pi n n'}
\end{align}
where 
\begin{align}
\boldsymbol{\pi}_{\boldsymbol{r}} \equiv        - {\rm i} \,  \hbar   \nabla_{\boldsymbol{r}}   +  \frac{\hbar    \boldsymbol{\sigma} \times \left[ \nabla_{\boldsymbol{r}} U_{{\rm n}, 0}(\boldsymbol{r})  \right]     }{4 m c^2}  \equiv \boldsymbol{p}_{\boldsymbol{r}} + \Delta \boldsymbol{p}^{\rm rel}_{\boldsymbol{r}} \tag{S7}
\label{general momentum}
\end{align}
is an operator in position space and a matrix in spin space. For practical use, Eq.~\eqref{pi n n'} can be rewritten as
\begin{align}
     \boldsymbol{\pi}_{n, n'}  =  \frac{(2 \pi)^3}{\Omega_{\rm cry}} \int d \boldsymbol{r}  \,   u^{\dagger}_{n, \boldsymbol{0}}(\boldsymbol{r})    \boldsymbol{\pi}_{\boldsymbol{r}} u_{n', \boldsymbol{0}}(\boldsymbol{r})  \,, \tag{S8}
\label{pi n n' practical}
\end{align}
where $\Omega_{\rm cry}$ is the crystal (i.e. the normalization) volume.

\subsection{Term due to $\hat{U}_{\rm ext}$}

Also the second term in the right-hand side of Eq.~(E9)
is evaluated using standard techniques \cite{VoonWillatzen, Luttinger55a}. It should be kept in mind, however, that the external potential must be expressed in curvilinear coordinates $\boldsymbol{r}$. Therefore, while its expression in Cartesian coordinates does not depend on $\varepsilon$, its expression in $\boldsymbol{r}$ coordinates acquires a dependence on $\varepsilon$ via the coordinate transformation $\boldsymbol{r}_{\rm C} \rightarrow \boldsymbol{r}$. In general, the external potential does not admit an expansion in powers of $\varepsilon$. This is not a problem, as it can be incorporated non-perturbatively into the Luttinger-Kohn equations, as long as it retains a slow spatial dependence with respect to the scale of a unit cell. Therefore,
\begin{align}
    \big( \chi_{n, \boldsymbol{k}} \big|      \hat{U}_{\rm ext}      \big| \chi_{n', \boldsymbol{k}'} \big)  = \int d \boldsymbol{r}   \,  \chi^{\dagger}_{n, \boldsymbol{k}}(\boldsymbol{r} ) \, U_{\rm ext}(\boldsymbol{r}) \, \chi_{n', \boldsymbol{k}'}(\boldsymbol{r})     \equiv  (2 \pi)^3 \sum_{\sigma} \sum_{\boldsymbol{G}} \sum_{\boldsymbol{G}'} \widetilde{u}_{n, \boldsymbol{0}}^{\sigma *}(\boldsymbol{G}) \, \widetilde{u}_{n', \boldsymbol{0}}^{\sigma}(\boldsymbol{G}') \, \widetilde{U}_{\rm ext}( \boldsymbol{k} - \boldsymbol{k}' + \boldsymbol{G} - \boldsymbol{G}' )       \,, \tag{S9}
    \label{V conf matrix element first step}
\end{align}
where
\begin{align}
\widetilde{U}_{\rm ext}(\boldsymbol{k}) \equiv \frac{1}{(2 \pi)^3} \int d \boldsymbol{r} \, {\rm e}^{- {\rm i} \boldsymbol{k} \cdot \boldsymbol{r}} \,  U_{\rm ext}(\boldsymbol{r})   \,. \tag{S10}
\end{align}
Assuming that the external potential is smooth over a lattice unit cell, it is posited that $\widetilde{U}_{\rm ext}$ in Eq.~\eqref{V conf matrix element first step} is not zero only if its argument lies inside the first Brillouin zone. Since $\boldsymbol{k}$ and $\boldsymbol{k}'$ are both inside the first Brillouin zone, the quantity $\boldsymbol{k} - \boldsymbol{k}' + \boldsymbol{G} - \boldsymbol{G}' $ satisfies the requirement only if $\boldsymbol{G} = \boldsymbol{G}'$ or if $\boldsymbol{G} - \boldsymbol{G}'$ is a nearest-neighbour of the origin in the reciprocal lattice. In the latter case, vectors $\boldsymbol{k}$ and $\boldsymbol{k}'$ can satisfy the constraint if they are close to opposite sides of the first Brillouin zone. Nevertheless, this case is usually neglected, and only $\boldsymbol{G} = \boldsymbol{G}'$ is considered. Under this approximation, the standard result is obtained: 
\begin{align}
    \big( \chi_{n, \boldsymbol{k}} \big|      \hat{U}_{\rm ext}      \big| \chi_{n', \boldsymbol{k}'} \big)   \approx   \delta_{n, n'} \, \widetilde{U}_{\rm ext}( \boldsymbol{k} - \boldsymbol{k}'  )       \,. \tag{S11}
    \label{V conf matrix element final}
\end{align}
Analogous approximations will be adopted in the remainder of the derivation, while dealing with the inhomogeneous-strain terms.

\subsection{Orthogonality correction}

The third term of the right-hand side of Eq.~(E9)
is the strain-dependent correction that ensures the orthonormality of the basis set. The corresponding Hamiltonian term is a slowly-varying function of position, formally analogous to an additional external potential. Therefore, the treatment of this term is analogous to that of $\hat{U}_{\rm ext}$ in the previous Subsection. Applying the same approximation, one obtains:
\begin{align}
   \big( \chi_{n, \boldsymbol{k}} \big|       \frac{\hbar^2}{4m} \left( \nabla^2 {\rm tr} \hat{\varepsilon} \right)     \big| \chi_{n', \boldsymbol{k}'} \big)  \approx - \delta_{n, n'}    \frac{  \hbar^2  }{4 m}       \left| \boldsymbol{k} - \boldsymbol{k}'  \right|^2         \widetilde{\varepsilon}^{\, \mu}_{\mu}(\boldsymbol{k} - \boldsymbol{k}')        \,. \tag{S12}
  \label{orthogonality correction - matrix element}
\end{align}

\subsection{Terms due to $\hat{H}_1$}

To evaluate the contributions to Eq.~(E9)
due to $\hat{H}_1$, it is convenient to split them into a nonrelativistic term and a spin-orbit term, as in Eq.~(E4).
The two terms will be considered separately.

\subsubsection{Nonrelativistic term}

The nonrelativistic term is given by the expression:
\begin{align}
    \big( \chi_{n, \boldsymbol{k}} \big|      \hat{H}_{1, {\rm nonrel}}     \big| \chi_{n', \boldsymbol{k}'} \big)  
    & =  \sum_{\sigma}  \int d \boldsymbol{r}  \Bigg\{   {\rm e}^{- {\rm i} \left( \boldsymbol{k} - \boldsymbol{k}' \right) \cdot \boldsymbol{r}}   u^{\sigma *}_{n, \boldsymbol{0}}(\boldsymbol{r} )   \,  \varepsilon^{\alpha}_{\beta} \left(\boldsymbol{r} \right)         U^{\beta}_{\alpha}(\boldsymbol{r} ) \,    u^{\sigma}_{n', \boldsymbol{0}}(\boldsymbol{r}) \nonumber \\
    & \quad - \frac{\hbar^2}{2 m}   \frac{ \partial  \left[ {\rm e}^{- {\rm i}   \boldsymbol{k}  \cdot \boldsymbol{r}}    u^{\sigma *}_{n, \boldsymbol{0}}(\boldsymbol{r}) \right] }{\partial r^{\mu}}       \left[       \varepsilon^{\nu}_{\mu}(\boldsymbol{r})    +    \varepsilon^{\mu}_{\nu}(\boldsymbol{r})        \right]  \frac{\partial \left[ u^{\sigma}_{n', \boldsymbol{0}}(\boldsymbol{r})  {\rm e}^{  {\rm i}   \boldsymbol{k}'   \cdot \boldsymbol{r}} \right]}{  \partial r^{\nu}}                \Bigg\}        \nonumber \\
    & = (2 \pi)^3 \sum_{\sigma}   \sum_{\boldsymbol{G}}   \widetilde{u}_{n, \boldsymbol{0}}^{\sigma *}(\boldsymbol{G})  \sum_{\boldsymbol{G}'}   \widetilde{u}_{n', \boldsymbol{0}}^{\sigma}(\boldsymbol{G}')     \Bigg\{              \sum_{\boldsymbol{G}''} \widetilde{\varepsilon}^{\, \mu}_{\nu}(\boldsymbol{k} - \boldsymbol{k}'  + \boldsymbol{G} - \boldsymbol{G}' - \boldsymbol{G}'' ) \, \widetilde{U}^{ \nu}_{\mu}(\boldsymbol{G}'' ) \nonumber \\
    & \quad - \frac{\hbar^2}{2 m}  \widetilde{\varepsilon}^{\, \mu}_{\nu}(\boldsymbol{k} - \boldsymbol{k}'   + \boldsymbol{G} - \boldsymbol{G}') \left[ \left( G^{\mu} + k^{\mu} \right)    \left( G'^{\nu} + k'^{\nu} \right)  
    + \left( G^{\nu} + k^{\nu} \right)    \left( G'^{\mu} + k'^{\mu} \right)  \right]                  \Bigg\}  \,, \tag{S13}
    \label{H epsilon NR matrix element second step}
\end{align}
where the Fourier transforms listed at the beginning of this Section have been used.

Since the strain tensor is assumed to have a slow spatial dependence, $\widetilde{\varepsilon}^{\, \mu}_{\nu}(\boldsymbol{q})$ is non-zero only if $\boldsymbol{q}$ belongs to the first Brillouin zone. The arguments of the Fourier transform of the strain tensor in Eq. \eqref{H epsilon NR matrix element second step} are $(\boldsymbol{k} - \boldsymbol{k}'   + \boldsymbol{G} - \boldsymbol{G}' - \boldsymbol{G}'')$ and $(\boldsymbol{k} - \boldsymbol{k}'   + \boldsymbol{G} - \boldsymbol{G}')$. In the first case, the requirement of slow spatial variation of the strain tensor imposes that $\boldsymbol{G} - \boldsymbol{G}' - \boldsymbol{G}''$ is either zero, or one of the nearest neighbours of the origin in the reciprocal space; in the second case, the same holds for $\boldsymbol{G} - \boldsymbol{G}'$.  

Along the lines of the approximation that is usually adopted for the slowly-varying confining potential (see the previous Subsection), we set to zero the combinations of reciprocal lattice vectors which are summed with $\boldsymbol{k} - \boldsymbol{k}'$ inside the arguments of the slowly-varying functions. Then, the following identities are introduced:
\begin{align}
    & \left( 2 \pi \right)^3      \sum_{\boldsymbol{G}}   \widetilde{u}_{n, \boldsymbol{0}}^{\dagger}(\boldsymbol{G}) \cdot    \widetilde{u}_{n', \boldsymbol{0}} (\boldsymbol{G}) = \frac{(2 \pi)^3}{\Omega_{\rm cry}}   \int d \boldsymbol{r} \, u^{\dagger}_{n, \boldsymbol{0}}(\boldsymbol{r}) \cdot   u_{n', \boldsymbol{0}}(\boldsymbol{r}) = \delta_{n, n'} \,, \nonumber \\
    & \left( 2 \pi \right)^3       \sum_{\boldsymbol{G}}   \widetilde{u}_{n, \boldsymbol{0}}^{\dagger}(\boldsymbol{G})      \cdot \widetilde{u}_{n', \boldsymbol{0}}(\boldsymbol{G}) \, G^{\nu} = \frac{(2 \pi)^3}{\Omega_{\rm cry}} \int d \boldsymbol{r} \, u^{\dagger}_{n, \boldsymbol{0}}(\boldsymbol{r}) \cdot \left( - {\rm i} \frac{\partial}{\partial r^{\nu} } \right)  u_{n', \boldsymbol{0}}(\boldsymbol{r})  \equiv \frac{1}{\hbar} \left( p_{\nu} \right)_{n,n'}  \,, \nonumber \\
    & \left( 2 \pi \right)^3     \sum_{\boldsymbol{G}}   \widetilde{u}_{n, \boldsymbol{0}}^{\dagger}(\boldsymbol{G})  \cdot   \widetilde{u}_{n', \boldsymbol{0}}(\boldsymbol{G}) \, G^{\nu} G^{\mu} = \frac{(2 \pi)^3}{\Omega_{\rm cry}} \int d \boldsymbol{r} \, u^{\dagger}_{n, \boldsymbol{0}}(\boldsymbol{r}) \cdot \left( - {\rm i} \frac{\partial}{\partial r^{\nu} } \right) \left( - {\rm i} \frac{\partial}{\partial r^{\mu} } \right)   u_{n', \boldsymbol{0}}(\boldsymbol{r})  \equiv \frac{1}{\hbar^2} \left(p_{ \mu} p_{\nu} \right)_{n,n'}  \,, \nonumber \\
    & \left( 2 \pi \right)^3    \sum_{\boldsymbol{G}} \sum_{\boldsymbol{G}'}   \widetilde{u}_{n, \boldsymbol{0}}^{\dagger}(\boldsymbol{G})  \cdot   \widetilde{u}_{n', \boldsymbol{0}}(\boldsymbol{G}')    \widetilde{U}_{ \mu}^{\nu}(\boldsymbol{G} - \boldsymbol{G}' )  = \frac{(2 \pi)^3}{\Omega_{\rm cry}} \int d \boldsymbol{r} \, U_{\mu}^{\nu}(\boldsymbol{r}) \, u^{\dagger}_{n, \boldsymbol{0}}(\boldsymbol{r}) \cdot    u_{n', \boldsymbol{0}}(\boldsymbol{r}) \equiv \left( U_{\mu}^{\nu} \right)_{n, n'} \,, \tag{S14}
\end{align}
where $\Omega_{\rm cry}$ is the crystal (i.e. the normalization) volume. 

In terms of these quantities, the result reads as
\begin{align}
     \big( \chi_{n, \boldsymbol{k}} \big|      \hat{H}_{1, {\rm nonrel}}     \big| \chi_{n', \boldsymbol{k}'} \big)      & \approx \widetilde{\varepsilon}^{\, \mu}_{\nu}(\boldsymbol{k} - \boldsymbol{k}')    \Bigg\{   \left( U_{\mu}^{\nu} \right)_{n, n'}   - \frac{\hbar^2}{2 m}     \Bigg[    \frac{2}{\hbar^2} \left(p_{ \mu} p_{\nu} \right)_{n,n'}  + \frac{1}{\hbar} \left( p_{\mu} \right)_{n,n'} \left( k^{\nu}  + k'^{\nu}  \right)    \nonumber \\
     & \quad +  \left(  k^{\mu} + k'^{\mu}\right) \frac{1}{\hbar} \left( p_{\nu} \right)_{n,n'} + \left( k^{\mu} k'^{\nu} + k'^{\mu} k^{\nu}   \right) \delta_{n, n'}      \Bigg]                  \Bigg\}   \,. \tag{S15}
    \label{H epsilon NR matrix element final, slow approximation}
\end{align}
The content of the curly braces in Eq.~\eqref{H epsilon NR matrix element final, slow approximation} can be expressed in terms of the quantities given in Eqs.~(13a-c) of the main text.

\subsubsection{Relativistic term}

The relativistic contribution is due to the operator
\begin{align}
    H_{1, {\rm so}} =   - \frac{ {\rm i} \hbar^2   }{8 m^2 c^2 }        \left(   \Sigma^{\nu} \overrightarrow{\partial}_{\nu}     -    \overleftarrow{\partial}_{\nu}    \Sigma^{\nu}  \right)   \,, \tag{S16}
\end{align}
where $\Sigma^{\nu}$ is defined in Eq.~(10).
It is convenient to elaborate this quantity as follows:
\begin{align}
    \Sigma^{\nu}(\boldsymbol{r}) & =  \left\{ \frac{\partial}{\partial r^{\mu}} \left[  \varepsilon^{\alpha}_{\beta}(\boldsymbol{r}) \,   U^{\beta}_{\alpha}(\boldsymbol{r})   \right]  \right\} \sigma^{\mu \nu}    +  \left(   \frac{\partial U_{{\rm n}, 0}(\boldsymbol{r}) }{\partial r^{\alpha}} \right)  \left[  \varepsilon^{\nu}_{\mu}(\boldsymbol{r})     \,     \sigma^{\mu \alpha} 
    -   \varepsilon^{\alpha}_{\mu}(\boldsymbol{r}) \,       \sigma^{\mu \nu} \right] \nonumber \\
    & =   {\rm i} \sum_{\boldsymbol{G}''} \int d \boldsymbol{q} \, {\rm e}^{{\rm i} \left( \boldsymbol{q} + \boldsymbol{G}'' \right) \cdot \boldsymbol{r}}  \left\{ \left[ \widetilde{U}^{\beta}_{\alpha}(\boldsymbol{G}'')    \, \widetilde{\varepsilon}^{\, \alpha}_{\beta}(\boldsymbol{q}) \, \left( q^{\mu} + G''^{\mu} \right)     
    -        G''^{\alpha} \widetilde{U}_{{\rm n}, 0}(\boldsymbol{G}'')  \,  \widetilde{\varepsilon}^{\, \alpha}_{\mu}(\boldsymbol{q})  \right]    \sigma^{\mu \nu}   
    +  G''^{\alpha} \widetilde{U}_{{\rm n}, 0}(\boldsymbol{G}'')  \,    \widetilde{\varepsilon}^{\, \nu}_{\mu}(\boldsymbol{q})     \sigma^{\mu \alpha}     \right\} \nonumber \\
    & \equiv   {\rm i} \sum_{\boldsymbol{G}''} \int d \boldsymbol{q} \, {\rm e}^{{\rm i} \left( \boldsymbol{q} + \boldsymbol{G}'' \right) \cdot \boldsymbol{r}}  \left[ S_{\mu}(\boldsymbol{q}, \boldsymbol{G}'' )    \,  \sigma^{\mu \nu}    + Z^{\nu}_{\mu \alpha}(\boldsymbol{q}, \boldsymbol{G}'' )  \, \sigma^{\mu \alpha}     \right]      \,. \tag{S17}
\end{align}
The matrix element is then:
\begin{align}
    \big( \chi_{n, \boldsymbol{k}} \big|    \hat{H}_{1, {\rm so}}    \big| \chi_{n', \boldsymbol{k}'} \big) & = - \frac{ {\rm i} \hbar^2   }{8 m^2 c^2 }    \int d \boldsymbol{r}    \left[  \chi^{\dagger}_{n, \boldsymbol{k}}(\boldsymbol{r})  \cdot  \Sigma^{\nu}(\boldsymbol{r}) \cdot \frac{ \partial  \chi_{n', \boldsymbol{k}'}(\boldsymbol{r}) }{\partial r^{\nu}}     -    \frac{\partial \chi^{\dagger}_{n, \boldsymbol{k}}(\boldsymbol{r})  }{\partial r^{\nu}}  \cdot   \Sigma^{\nu}(\boldsymbol{r}) \cdot \chi_{n', \boldsymbol{k}'}(\boldsymbol{r}) \right]  \nonumber \\
    & =  {\rm i} \frac{   \hbar^2   }{8 m^2 c^2 }  (2 \pi)^3 \sum_{\boldsymbol{G}, \boldsymbol{G}', \boldsymbol{G}'' } \widetilde{u}^{\dagger}_{n, \boldsymbol{0} }(\boldsymbol{G})  \cdot  \sigma^{\mu \nu} \cdot \widetilde{u}_{n', \boldsymbol{0}}(\boldsymbol{G}')   \Big\{  \left( k^{\nu} + G^{\nu} + k'^{\nu} + G'^{\nu} \right)      S_{\mu} \left( \boldsymbol{q}, \boldsymbol{G}'' \right)       \nonumber \\
    & \quad + \left( k^{\alpha} + G^{\alpha} + k'^{\alpha} + G'^{\alpha} \right)      Z^{\alpha}_{\mu \nu} \left( \boldsymbol{q}, \boldsymbol{G}'' \right)  \Big\} \Big|_{\boldsymbol{q} = \boldsymbol{k} + \boldsymbol{G} - \boldsymbol{k}' - \boldsymbol{G}' - \boldsymbol{G}''}  \,. \tag{S18}
\end{align}

Consistently with the approximation that was already discussed for the previous terms, here $\boldsymbol{G}'' = \boldsymbol{G} - \boldsymbol{G}'$ should be substituted in the whole expression. In particular,
\begin{align}
    S_{\mu} \left( \boldsymbol{q}, \boldsymbol{G}'' \right) \Big|_{\boldsymbol{q} = \boldsymbol{k} + \boldsymbol{G} - \boldsymbol{k}' - \boldsymbol{G}' - \boldsymbol{G}''} & \approx \delta_{\boldsymbol{G}'' , \boldsymbol{G} - \boldsymbol{G}'} \Big[ \widetilde{U}^{\beta}_{\alpha}(\boldsymbol{G} - \boldsymbol{G}')    \, \widetilde{\varepsilon}^{\, \alpha}_{\beta}(\boldsymbol{k}   - \boldsymbol{k}'  ) \, \left( k^{\mu} + G^{\mu} - k'^{\mu} - G'^{\mu} \right)   \nonumber \\
    & \quad
    -      \left(   G^{\alpha} -  G'^{\alpha} \right) \widetilde{U}_{{\rm n}, 0}(\boldsymbol{G} - \boldsymbol{G}')    \widetilde{\varepsilon}^{\, \alpha}_{\mu}(\boldsymbol{k}   - \boldsymbol{k}' )  \Big]  \,, \tag{S19}
\end{align}
and
\begin{align}
    Z^{\alpha}_{\mu \nu}(\boldsymbol{q}, \boldsymbol{G}'' ) \Big|_{\boldsymbol{q} = \boldsymbol{k} + \boldsymbol{G} - \boldsymbol{k}' - \boldsymbol{G}' - \boldsymbol{G}''}   \approx     \delta_{\boldsymbol{G}'' , \boldsymbol{G} - \boldsymbol{G}'}  \left( G^{\alpha} - G'^{\alpha} \right) \widetilde{U}_{{\rm n}, 0}(\boldsymbol{G} - \boldsymbol{G}')      \widetilde{\varepsilon}^{\, \nu}_{\mu}(\boldsymbol{k} - \boldsymbol{k}')    \,. \tag{S20}
\end{align}

Finally, the following identities hold for any function $U(\boldsymbol{r})$ having the same periodicity as the lattice:
\begin{align}
    & \left( 2 \pi \right)^3     \sum_{\boldsymbol{G} , \boldsymbol{G}'}   \widetilde{u}_{n, \boldsymbol{0}}^{\dagger}(\boldsymbol{G})  \cdot \sigma^{\mu \nu} \cdot   \widetilde{u}_{n', \boldsymbol{0}}(\boldsymbol{G}')  \,  \widetilde{U}(\boldsymbol{G} - \boldsymbol{G}' )  =  \frac{(2 \pi)^3}{\Omega_{\rm cry}} \int d \boldsymbol{r} \, u^{\dagger}_{n, \boldsymbol{0}}(\boldsymbol{r})  \cdot \sigma^{\mu \nu} \cdot   u_{n', \boldsymbol{0}}(\boldsymbol{r}) \, U(\boldsymbol{r}) \equiv \left( \sigma^{\mu \nu} U \right)_{n, n'} \,, \nonumber \\ 
    & \left( 2 \pi \right)^3    \sum_{\boldsymbol{G} , \boldsymbol{G}' }   \widetilde{u}_{n, \boldsymbol{0}}^{\dagger}(\boldsymbol{G})  \cdot \sigma^{\mu \nu}  \cdot  \widetilde{u}_{n', \boldsymbol{0}}(\boldsymbol{G}') \,  \widetilde{U}(\boldsymbol{G} - \boldsymbol{G}' ) \, \left( G^{\alpha} - G'^{\alpha} \right) = \frac{(2 \pi)^3}{\Omega_{\rm cry}}  \int d \boldsymbol{r} \, u^{\dagger}_{n, \boldsymbol{0}}(\boldsymbol{r}) \cdot \sigma^{\mu \nu} \cdot u_{n', \boldsymbol{0}}(\boldsymbol{r}) \, \left( - {\rm i} \frac{\partial U(\boldsymbol{r})}{\partial r^{\alpha} } \right)      \nonumber \\
        &  \equiv \frac{1}{\hbar} \left[ \sigma^{\mu \nu} \left( p_{\alpha} U \right) \right]_{n,n'}  \,, \nonumber \\ 
    & \left( 2 \pi \right)^3    \sum_{\boldsymbol{G}, \boldsymbol{G}'}   \widetilde{u}_{n, \boldsymbol{0}}^{\dagger}(\boldsymbol{G})  \cdot \sigma^{\mu \nu}  \cdot  \widetilde{u}_{n', \boldsymbol{0}}(\boldsymbol{G}') \,  \widetilde{U}(\boldsymbol{G} - \boldsymbol{G}' ) \,   G^{\alpha}   =  \frac{(2 \pi)^3}{\Omega_{\rm cry}} \int d \boldsymbol{r} \, \left( {\rm i} \frac{ \partial u^{\dagger}_{n, \boldsymbol{0}}(\boldsymbol{r}) }{\partial r^{\alpha}} \right) \cdot \sigma^{\mu \nu} \cdot      u_{n', \boldsymbol{0}}(\boldsymbol{r})   U(\boldsymbol{r})  \nonumber \\
        & \equiv \frac{1}{\hbar} \left[ \overleftarrow{p}_{\alpha} \left( \sigma^{\mu \nu} U  \right)      \right]_{n,n'}  \,,   \nonumber \\
    & \left( 2 \pi \right)^3    \sum_{\boldsymbol{G}, \boldsymbol{G}'}   \widetilde{u}_{n, \boldsymbol{0}}^{\dagger}(\boldsymbol{G})  \cdot \sigma^{\mu \nu}  \cdot  \widetilde{u}_{n', \boldsymbol{0}}(\boldsymbol{G}') \,  \widetilde{U}(\boldsymbol{G} - \boldsymbol{G}' ) \,   G'^{\alpha}   =  \frac{(2 \pi)^3}{\Omega_{\rm cry}}   \int d \boldsymbol{r} \, u^{\dagger}_{n, \boldsymbol{0}}(\boldsymbol{r}) \cdot \sigma^{\mu \nu} \cdot    \left( - {\rm i} \frac{\partial u_{n', \boldsymbol{0}}(\boldsymbol{r}) }{\partial r^{\alpha} } \right) U(\boldsymbol{r})  \nonumber \\
        & \equiv \frac{1}{\hbar} \left[   \left( \sigma^{\mu \nu} U  \right) \overrightarrow{p}_{\alpha}    \right]_{n,n'}  \,, \nonumber \\  
    & \left( 2 \pi \right)^3    \sum_{\boldsymbol{G}, \boldsymbol{G}'}   \widetilde{u}_{n, \boldsymbol{0}}^{\dagger}(\boldsymbol{G})  \cdot \sigma^{\mu \nu}  \cdot  \widetilde{u}_{n', \boldsymbol{0}}(\boldsymbol{G}') \,  \widetilde{U}(\boldsymbol{G} - \boldsymbol{G}' ) \,   G^{\alpha} G'^{\beta}   =   \frac{(2 \pi)^3}{\Omega_{\rm cry}}  \int d \boldsymbol{r} \, \left( {\rm i} \frac{ \partial u^{\dagger}_{n, \boldsymbol{0}}(\boldsymbol{r}) }{\partial r^{\alpha}} \right)  \cdot \sigma^{\mu \nu} \cdot    \left( - {\rm i} \frac{\partial u_{n', \boldsymbol{0}}(\boldsymbol{r}) }{\partial r^{\beta} } \right) U(\boldsymbol{r})     \nonumber \\
        &  \equiv \frac{1}{\hbar^2} \left[  \overleftarrow{p}_{\alpha}  \left( \sigma^{\mu \nu} U  \right) \overrightarrow{p}_{\beta}    \right]_{n,n'}  \,, \nonumber \\  
    & \left( 2 \pi \right)^3    \sum_{\boldsymbol{G}, \boldsymbol{G}'}   \widetilde{u}_{n, \boldsymbol{0}}^{\dagger}(\boldsymbol{G})  \cdot \sigma^{\mu \nu}  \cdot  \widetilde{u}_{n', \boldsymbol{0}}(\boldsymbol{G}') \,  \widetilde{U}(\boldsymbol{G} - \boldsymbol{G}' ) \, \left( G^{\nu} + G'^{\nu} \right) \left( G^{\mu} - G'^{\mu} \right)     \equiv \frac{1}{\hbar^2} \left[ \overleftarrow{p}_{\nu} \left( \sigma^{\mu \nu} p_{\mu} U  \right)   +  \left( \sigma^{\mu \nu} p_{\mu} U  \right) \overrightarrow{p}_{\nu}    \right]_{n,n'}  \,. \tag{S21}
\end{align}

The result is
\begin{align}
    \big( \chi_{n, \boldsymbol{k}} \big|    \hat{H}_{1, {\rm so}}    \big| \chi_{n', \boldsymbol{k}'} \big)  
    & \approx   {\rm i} \frac{   \hbar^2   }{8 m^2 c^2 }  \widetilde{\varepsilon}^{\, \alpha}_{\beta}(\boldsymbol{k}   - \boldsymbol{k}'  )    \Bigg\{      2 k^{\mu} k'^{\nu}  \left(  \sigma^{\mu \nu}    U^{\beta}_{\alpha} \right)_{n, n'}    +  2  k^{\mu}   \,  \frac{1}{\hbar} \left[   \left( \sigma^{\mu \nu} U^{\beta}_{\alpha}  \right) \overrightarrow{p}_{\nu}    \right]_{n,n'}  + 2 k'^{\nu}   \,    \frac{1}{\hbar} \left[ \overleftarrow{p}_{\mu} \left( \sigma^{\mu \nu} U^{\beta}_{\alpha}  \right)      \right]_{n,n'}     \nonumber \\
    & \quad  +   2 \frac{1}{\hbar^2} \left[  \overleftarrow{p}_{\mu}  \left( \sigma^{\mu \nu} U^{\beta}_{\alpha}  \right) \overrightarrow{p}_{\nu}    \right]_{n,n'}  
    +   \left( k^{\nu}  + k'^{\nu}   \right) \,    \frac{1}{\hbar} \left[ \sigma^{\beta \alpha} \left( p_{\nu} U_{{\rm n}, 0} \right) - \sigma^{\beta \nu} \left( p_{\alpha} U_{{\rm n}, 0} \right) \right]_{n,n'}     \nonumber \\
    & \quad +          \frac{1}{\hbar^2} \left[ \overleftarrow{p}_{\nu} \left( \sigma^{\beta \alpha} p_{\nu} U_{{\rm n}, 0} \right)  - \overleftarrow{p}_{\nu} \left( \sigma^{\beta \nu} p_{\alpha} U_{{\rm n}, 0} \right)     +  \left( \sigma^{\beta \alpha} p_{\nu} U_{{\rm n}, 0}  \right) \overrightarrow{p}_{\nu}   - \left( \sigma^{\beta \nu} p_{\alpha} U_{{\rm n}, 0}  \right) \overrightarrow{p}_{\nu}  \right]_{n,n'}   \Bigg\}    \,. \tag{S22}
\end{align}

\subsection{Total matrix element}

Combining the terms derived above, one can express Eq.~(E9)
as
\begin{align}
    \big< \overline{\chi}_{n, \boldsymbol{k}} \big|    \hat{H}    \big| \overline{\chi}_{n', \boldsymbol{k}'} \big> \approx  \big< \overline{\chi}_{n, \boldsymbol{k}} \big|    \hat{H}^{(0)}    \big| \overline{\chi}_{n', \boldsymbol{k}'} \big> 
    + \big< \overline{\chi}_{n, \boldsymbol{k}} \big|    \hat{H}^{(1)}    \big| \overline{\chi}_{n', \boldsymbol{k}'} \big>     \,, \tag{S23}
    \label{LK total}
\end{align}
where
\begin{align}
    \big< \overline{\chi}_{n, \boldsymbol{k}} \big|    \hat{H}^{(0)}    \big| \overline{\chi}_{n', \boldsymbol{k}'} \big> & \equiv \big(  \chi_{n, \boldsymbol{k}} \big|  \hat{H}_0 \big|   \chi_{n', \boldsymbol{k}'} \big)         + \big( \chi_{n, \boldsymbol{k}} \big|      \hat{U}_{\rm ext}    \big| \chi_{n', \boldsymbol{k}'} \big)  \nonumber \\
    & = \delta(\boldsymbol{k} - \boldsymbol{k}') \left\{ \delta_{n, n'}  \, \left[ E_n(\boldsymbol{0}) + \frac{ \hbar^2 \boldsymbol{k}^2 }{2 m} \right]   + \frac{\hbar }{m} \boldsymbol{k} \cdot \boldsymbol{\pi}_{n, n'}  \right\} + \delta_{n, n'} \, \widetilde{U}_{\rm ext}( \boldsymbol{k} - \boldsymbol{k}'  )  \tag{S24}
    \label{LK H0}
\end{align}
is formally analogous to the standard $\boldsymbol{k} \cdot \boldsymbol{p}$ Hamiltonian (but expressed in curvilinear coordinates), and 
\begin{align}
    \big< \overline{\chi}_{n, \boldsymbol{k}} \big|    \hat{H}^{(1)}    \big| \overline{\chi}_{n', \boldsymbol{k}'} \big> & \equiv \frac{  \hbar^2 }{4 m}  \big(  \chi_{n, \boldsymbol{k}} \big| \left( \nabla^2 {\rm tr} \varepsilon \right) \big|   \chi_{n', \boldsymbol{k}'} \big)  + \big( \chi_{n, \boldsymbol{k}} \big|      \hat{H}_1    \big| \chi_{n', \boldsymbol{k}'} \big) \nonumber \\
    & = - \frac{  \hbar^2  }{4 m}       \left| \boldsymbol{k} - \boldsymbol{k}' \right|^2         \widetilde{\varepsilon}^{\, \mu}_{\mu}(\boldsymbol{k} - \boldsymbol{k}') \, \delta_{n,n'} 
 + \widetilde{\varepsilon}^{\, \mu}_{\nu}(\boldsymbol{k} - \boldsymbol{k}')    \left(  \mathcal{D}_{\mu}^{\nu}        + k^{\alpha}  \mathcal{L}_{\alpha; \mu}^{\nu}   
+ k'^{\alpha} \mathcal{L}_{\alpha; \mu}^{*\nu} + k^{\alpha} k'^{\beta}  \mathcal{Q}_{\alpha \beta; \mu}^{\nu} \right)_{n, n'}   \tag{S25}
\label{LK H1}
\end{align}
is the correction due to strain, as reported in Eq.~(12) of the main text, where the deformation potentials are defined in Eqs.~(13)-(15).

\section*{II. Manifold decoupling}

We here apply L\"owdin partitioning in order to decouple a low-energy manifold of bands, with $n \in \lbrace 1, 2, \ldots, N \rbrace$, from the higher (remote) bands with $n > N$. Using the notation of Ref.~\cite{Winkler}, the Hamiltonian is written as
\begin{align}
    \hat{H} \equiv \hat{H}^{(0)} + \hat{H}^{(1)} + \hat{H}^{(2)} \,, \tag{S26}
\end{align}
where $\hat{H}^{(0)}$ is diagonal in the band and crystal-momentum indices, $\hat{H}^{(1)}$ contains all intra-manifold terms, and $\hat{H}^{(2)}$ contains the inter-manifold terms. In the case at hand, using Eqs.~\eqref{LK total}-\eqref{LK H1}, the three parts are written as
\begin{align}
     \hat{H}^{(0)}   \equiv \sum_{n  }  \int_{\rm 1 BZ} d \boldsymbol{k}      \left[ E_n(\boldsymbol{0}) + \frac{ \hbar^2 \boldsymbol{k}^2 }{2 m} \right]   \big| \overline{\chi}_{n, \boldsymbol{k}} \big>   \big< \overline{\chi}_{n, \boldsymbol{k}} \big|  \,,  \tag{S27}
     \label{H0 manifold}
\end{align}
\begin{align}
     \hat{H}^{(1)} & \equiv \left( \sum_{n \leq N} \sum_{n' \leq N}  + \sum_{n > N} \sum_{n' > N} \right) \int_{\rm 1 BZ} d \boldsymbol{k}  \frac{\hbar }{m} \boldsymbol{k} \cdot \boldsymbol{\pi}_{n, n'}     \big| \overline{\chi}_{n, \boldsymbol{k}} \big>      \big< \overline{\chi}_{n', \boldsymbol{k}} \big|  \nonumber \\
     & \quad + \sum_n  \int_{\rm 1BZ} d \boldsymbol{k} \int_{\rm 1BZ} d \boldsymbol{k}'   \left[ \widetilde{U}_{\rm ext}( \boldsymbol{k} - \boldsymbol{k}'  )  -    \frac{  \hbar^2  }{4 m}       \left| \boldsymbol{k} - \boldsymbol{k}'  \right|^2         \widetilde{\varepsilon}^{\, \mu}_{\mu}(\boldsymbol{k} - \boldsymbol{k}') \right] \big| \overline{\chi}_{n, \boldsymbol{k}} \big> \big< \overline{\chi}_{n, \boldsymbol{k}'} \big| \nonumber \\
     & \quad + \left( \sum_{n \leq N} \sum_{n' \leq N}  + \sum_{n > N} \sum_{n' > N} \right)   \int_{\rm 1BZ} d \boldsymbol{k} \int_{\rm 1BZ} d \boldsymbol{k}' \widetilde{\varepsilon}^{\, \mu}_{\nu}(\boldsymbol{k} - \boldsymbol{k}')    \left[ X^{\nu}_{\mu}(\boldsymbol{k}, \boldsymbol{k}') \right]_{n, n'}   \big| \overline{\chi}_{n, \boldsymbol{k}} \big> \big< \overline{\chi}_{n', \boldsymbol{k}'} \big| \,, \tag{S28}
     \label{H1 manifold}
\end{align}
\begin{align}
    \hat{H}^{(2)}   \equiv  \left( \sum_{n \leq N} \sum_{n' > N} + \sum_{n > N} \sum_{n' \leq N} \right) \! \int_{\rm 1BZ} d \boldsymbol{k} \int_{\rm 1BZ} d \boldsymbol{k}' \left\{ \delta(\boldsymbol{k} - \boldsymbol{k}') \frac{\hbar }{m} \boldsymbol{k} \cdot \boldsymbol{\pi}_{n, n'} + \widetilde{\varepsilon}^{\, \mu}_{\nu}(\boldsymbol{k} - \boldsymbol{k}')    \left[ X^{\nu}_{\mu}(\boldsymbol{k}, \boldsymbol{k}') \right]_{n, n'}  \right\}  \big| \overline{\chi}_{n, \boldsymbol{k}} \big> \big< \overline{\chi}_{n', \boldsymbol{k}'} \big|  , \tag{S29}
    \label{H2 manifold}
\end{align}
where
\begin{align} 
       \left[ X^{\nu}_{\mu}(\boldsymbol{k}, \boldsymbol{k}') \right]_{n, n'} = \left( \mathcal{D}_{\mu}^{\nu} \right)_{n, n'}       + k^{\alpha} \left( \mathcal{L}_{\alpha; \mu}^{\nu} \right)_{n, n'}   + k'^{\alpha} \left( \mathcal{L}_{\alpha; \mu}^{*\nu} \right)_{n', n} + k^{\alpha} k'^{\beta}  \left( \mathcal{Q}_{\alpha \beta; \mu}^{\nu} \right)_{n, n'}       \,. \tag{S30}
\end{align}

A canonical transformation
\begin{align}
    \hat{\mathcal{H}} = {\rm e}^{- \hat{S}} \hat{H} {\rm e}^{\hat{S}} \,, \quad \big| \phi \big> = {\rm e}^{- \hat{S}} \big| \psi \big> \,, \tag{S31}
\end{align}
is applied to block-diagonalize $\hat{H}^{(2)}$, while preserving the already block-diagonal form of $\hat{H}^{(0)} + \hat{H}^{(1)}$. Following the procedure outlined in Ref.~\cite{Winkler}, one writes
\begin{align}
    \hat{\mathcal{H}} = \hat{\mathcal{H}}_{\rm diag} + \hat{\mathcal{H}}_{\rm nondiag} \,, \tag{S32}
\end{align}
\begin{align}
    \hat{\mathcal{H}}_{\rm diag} & = \sum_{j = 0}^{\infty} \frac{1}{(2 j)!} \left[ \hat{H}^{(0)} + \hat{H}^{(1)} \,,\, \hat{S} \right]^{(2j)} 
    + \sum_{j = 0}^{\infty} \frac{1}{(2 j + 1)!} \left[ \hat{H}^{(2)} \,,\, \hat{S} \right]^{(2j + 1)} \nonumber \\ 
    & =   \hat{H}^{(0)} + \hat{H}^{(1)} +     \left[ \hat{H}^{(2)} \,,\, \hat{S} \right]    +   \frac{1}{2} \left[ \left[ \hat{H}^{(0)} + \hat{H}^{(1)} \,,\, \hat{S} \right] \,,\,  \hat{S} \right] 
      + \ldots   \,, \tag{S33}
\end{align}
\begin{align}
    \hat{\mathcal{H}}_{\rm nondiag} & = \sum_{j = 0}^{\infty} \frac{1}{(2 j + 1)!} \left[ \hat{H}^{(0)} + \hat{H}^{(1)} \,,\, \hat{S} \right]^{(2j + 1)} 
    + \sum_{j = 0}^{\infty} \frac{1}{(2 j)!} \left[ \hat{H}^{(2)} \,,\, \hat{S} \right]^{(2j)} \nonumber \\
    & =  \hat{H}^{(2)}   +    \left[ \hat{H}^{(0)} + \hat{H}^{(1)} \,,\, \hat{S} \right]  
    +      \frac{1}{2} \left[ \left[ \hat{H}^{(2)} \,,\, \hat{S} \right] \,,\, \hat{S} \right]  + \ldots \,. \tag{S34}
\end{align}
The operator $\hat{S}$ is chosen so that $\hat{\mathcal{H}}_{\rm nondiag} \approx 0$. Expanding $\hat{S} = \hat{S}^{(1)} + \hat{S}^{(2)} + \ldots$, this condition is satisfied by imposing  
\begin{align}
    \left[ \hat{H}^{(0)} \,,\, \hat{S}^{(1)}\right] = - \hat{H}^{(2)} \,, \quad \left[ \hat{H}^{(0)} \,,\, \hat{S}^{(2)}\right] = - \left[ \hat{H}^{(1)}, \hat{S}^{(1)} \right] \,, \ldots \quad . \tag{S35}
    \label{conditions S1 S2}
\end{align}
The resulting first term of the expansion is
\begin{align}
    \hat{S}^{(1)}  =   - \left( \sum_{n \leq N} \sum_{n' > N} + \sum_{n > N} \sum_{n' \leq N} \right)  \int_{\rm 1BZ} d \boldsymbol{k} \int_{\rm 1BZ} d \boldsymbol{k}' \frac{ \delta(\boldsymbol{k} - \boldsymbol{k}') \frac{\hbar }{m} \boldsymbol{k} \cdot \boldsymbol{\pi}_{n, n'} + \widetilde{\varepsilon}^{\, \mu}_{\nu}(\boldsymbol{k} - \boldsymbol{k}')    \left[ X^{\nu}_{\mu}(\boldsymbol{k}, \boldsymbol{k}') \right]_{n, n'}  }{   E_n(\boldsymbol{0}) + \frac{ \hbar^2 \boldsymbol{k}^2 }{2 m} - E_{n'}(\boldsymbol{0}) - \frac{ \hbar^2 \boldsymbol{k}'^2 }{2 m}    } \big| \overline{\chi}_{n, \boldsymbol{k}} \big> \big< \overline{\chi}_{n', \boldsymbol{k}'} \big| \,. \tag{S36}
    \label{S1 solved}
\end{align}
The second term, $S^{(2)}$, displays an additional large denominator with respect to $S^{(1)}$, so it is much smaller and it will be neglected here.

The effective Hamiltonian after the canonical transformation is then
\begin{align}
    \hat{\mathcal{H}} \approx \hat{H}^{(0)} + \hat{H}^{(1)} + \left[ \hat{H}^{(2)} \,,\, \hat{S}^{(1)} \right] \,, \tag{S37}
\end{align}
where terms of order $\propto \varepsilon^2$ must be discarded from the expression of $\left[ \hat{H}^{(2)} \,,\, \hat{S}^{(1)} \right]$, since the present theory is accurate only up to the first order in the strain tensor. The resulting Hamiltonian restricted to the $n \leq N$ manifold is:
\begin{align}
    \hat{\mathcal{H}}^{(N)} & \approx  \sum_{n \leq N }  \int_{\rm 1 BZ} d \boldsymbol{k}      \left[ E_n(\boldsymbol{0}) + \frac{ \hbar^2 \boldsymbol{k}^2 }{2 m} \right]   \big| \overline{\chi}_{n, \boldsymbol{k}} \big>   \big< \overline{\chi}_{n, \boldsymbol{k}} \big|   +   \sum_{n \leq N} \sum_{n' \leq N}   \int_{\rm 1 BZ} d \boldsymbol{k}  \frac{\hbar }{m} \boldsymbol{k} \cdot \boldsymbol{\pi}_{n, n'}     \big| \overline{\chi}_{n, \boldsymbol{k}} \big>      \big< \overline{\chi}_{n', \boldsymbol{k}} \big|  \nonumber \\
         & \quad + \sum_{n \leq N} \sum_{n' \leq N} \sum_{n'' > N}         
    \int_{\rm 1BZ} d \boldsymbol{k}            \frac{\hbar }{m} \boldsymbol{k} \cdot \boldsymbol{\pi}_{n, n''}    \frac{\hbar }{m} \boldsymbol{k} \cdot \boldsymbol{\pi}_{n'', n'}   \left( \frac{ 1  }{   E_{n}(\boldsymbol{0})   - E_{n''}(\boldsymbol{0})     } +  \frac{  1   }{ E_{n'}(\boldsymbol{0})    - E_{n''}(\boldsymbol{0})      }          \right)   \big| \overline{\chi}_{n, \boldsymbol{k}} \big>  \big< \overline{\chi}_{n', \boldsymbol{k}} \big| \nonumber \\
     & \quad + \sum_{n \leq N}  \int_{\rm 1BZ} d \boldsymbol{k} \int_{\rm 1BZ} d \boldsymbol{k}'   \left[ \widetilde{U}_{\rm ext}( \boldsymbol{k} - \boldsymbol{k}'  )  -    \frac{  \hbar^2  }{4 m}       \left| \boldsymbol{k} - \boldsymbol{k}'  \right|^2         \widetilde{\varepsilon}^{\, \mu}_{\mu}(\boldsymbol{k} - \boldsymbol{k}') \right] \big| \overline{\chi}_{n, \boldsymbol{k}} \big> \big< \overline{\chi}_{n, \boldsymbol{k}'} \big| \nonumber \\
     & \quad +   \sum_{n \leq N} \sum_{n' \leq N}      \int_{\rm 1BZ} d \boldsymbol{k} \int_{\rm 1BZ} d \boldsymbol{k}' \widetilde{\varepsilon}^{\, \mu}_{\nu}(\boldsymbol{k} - \boldsymbol{k}')    \left[ X^{\nu}_{\mu}(\boldsymbol{k}, \boldsymbol{k}') \right]_{n, n'}   \big| \overline{\chi}_{n, \boldsymbol{k}} \big> \big< \overline{\chi}_{n', \boldsymbol{k}'} \big| \nonumber \\
     & \quad   + \sum_{n \leq N} \sum_{n' \leq N} \sum_{n'' > N}         
    \int_{\rm 1BZ} d \boldsymbol{k} \int_{\rm 1BZ} d \boldsymbol{k}'   \widetilde{\varepsilon}^{\, \mu}_{\nu}(\boldsymbol{k} - \boldsymbol{k}') \big| \overline{\chi}_{n, \boldsymbol{k}} \big>  \big< \overline{\chi}_{n', \boldsymbol{k}'} \big|    \nonumber \\ 
    & \quad   \times    \Bigg[   \frac{\hbar }{m} \boldsymbol{k} \cdot \boldsymbol{\pi}_{n, n''}           \left[ X^{\nu}_{\mu}(\boldsymbol{k}, \boldsymbol{k}') \right]_{n'', n'}      \left( \frac{ 1  }{   E_{n}(\boldsymbol{0})   - E_{n''}(\boldsymbol{0})     } +  \frac{  1   }{ E_{n'}(\boldsymbol{0}) + \frac{ \hbar^2 \boldsymbol{k}'^2 }{2 m}  - E_{n''}(\boldsymbol{0}) - \frac{ \hbar^2 \boldsymbol{k}^2 }{2 m}     }          \right)     \nonumber \\
     & \quad \quad       +         \left[ X^{\nu}_{\mu}(\boldsymbol{k}, \boldsymbol{k}') \right]_{n, n''}     \frac{\hbar }{m} \boldsymbol{k}' \cdot \boldsymbol{\pi}_{n'', n'}      \left( \frac{ 1  }{   E_{n}(\boldsymbol{0}) + \frac{ \hbar^2 \boldsymbol{k}^2 }{2 m} - E_{n''}(\boldsymbol{0}) - \frac{ \hbar^2 \boldsymbol{k}'^2 }{2 m}    } +  \frac{  1   }{ E_{n'}(\boldsymbol{0})  - E_{n''}(\boldsymbol{0})     }          \right)  \Bigg]    \,. \tag{S38}
     \label{Hamiltonian N block}
\end{align}
The last three lines represent a small contribution with respect to the dominant, strain-independent one, and they will be neglected here. Under this approximation, the matrix elements of Eq.~\eqref{Hamiltonian N block} are written as 
\begin{align}
 \big< \overline{\chi}_{n, \boldsymbol{k}} \big| 
 \hat{\mathcal{H}} \big| \overline{\chi}_{n', \boldsymbol{k}'} \big>     \approx  \big< \overline{\chi}_{n, \boldsymbol{k}} \big|      \hat{H}       \big| \overline{\chi}_{n', \boldsymbol{k}'} \big> + \delta(\boldsymbol{k} - \boldsymbol{k}') \frac{\hbar^2 \Pi^{\alpha \beta}_{n, n'} }{m^2} k^{\alpha}  k^{\beta}       \,, \tag{S39}
     \label{Hamiltonian N block matrelem}
\end{align} 
where $n , n' \leq N$, and
\begin{align}
    \Pi^{\alpha \beta}_{n, n'} & \equiv \sum_{n'' > N}     \left( \frac{  \pi^{\alpha}_{n, n''}     \pi^{\beta}_{n'', n'} }{   E_{n}(\boldsymbol{0})   - E_{n''}(\boldsymbol{0})     }  +  \frac{   \pi^{\alpha}_{n, n''}     \pi^{\beta}_{n'', n'}   }{ E_{n'}(\boldsymbol{0})    - E_{n''}(\boldsymbol{0})      }          \right)  \,. \tag{S40}
\end{align}

\end{widetext}


\begin{thebibliography}{50}
\bibitem [{let()}]{letmeexplain}%
  \BibitemOpen
  \href@noop {} {}\bibinfo {note} {The reason for this is the fact that, for nonzero $\varepsilon$, both the $\varepsilon$-independent part and the $\varepsilon$-dependent part of $H$ are not Hermitian operators. Their sum, instead, is Hermitian.}\BibitemShut {Stop}%
\end{thebibliography}
\end{document}